\documentclass[jcp, twocolumn, floatfix, superscriptaddress, reprint, 10pt]{revtex4-1}
\usepackage[english]{babel}
\usepackage[utf8]{inputenc}
\usepackage[paperwidth=210mm,paperheight=297mm,centering,hmargin=2cm,vmargin=2.5cm]{geometry}
\usepackage[utf8]{inputenc}
\usepackage{verbatim}
\usepackage{bm}
\usepackage{amsmath}
\usepackage{natbib}
\usepackage{pdfpages}
\usepackage{graphics}
\usepackage{float}
\usepackage{hyperref}
\usepackage[capitalize]{cleveref}
\usepackage{multirow}
\usepackage{makecell}
\usepackage{outlines}
\usepackage{siunitx}
\usepackage{mhchem}
\usepackage{caption}
\usepackage{subcaption}
\usepackage{xcolor}
\usepackage{ulem}
\ifdefined\DIRACREP
\else
\newcommand{\DIRACREP}{}

\usepackage{physics}
\usepackage{xparse}
\ifdefined\COSMOMATHS
\else
\newcommand{\COSMOMATHS}{}

\newcommand{\mbf}[1]{\ensuremath{\mathbf{#1}}}

\newcommand{\D}[1]{\operatorname{d}{\!#1}\,}

\fi %

\NewDocumentCommand{\rep}{s d<| d|>}{%
\IfBooleanTF{#1}{
   \IfValueTF{#2}{
       \IfValueTF{#3}{\braket{#2}{#3}}{\bra{#2}}
       }{
       \IfValueTF{#3}{\ket{#3}}{}
       }
   }{
   \IfValueTF{#2}{
       \IfValueTF{#3}{\braket*{#2}{#3}}{\bra*{#2}}
       }{
       \IfValueTF{#3}{\ket*{#3}}{}
       }
   }
}

\NewDocumentCommand{\rbra}{sm}{\IfBooleanTF{#1}{\rep*<#2|}{\rep<#2|}}
\NewDocumentCommand{\rket}{sm}{\IfBooleanTF{#1}{\rep*|#2>}{\rep|#2>}}
\NewDocumentCommand{\rbraket}{smom}{
    \IfBooleanTF{#1}{
        \IfNoValueTF{#3}{\rep*<#2||#4>}{\rep*<#2|#3\rep*|#4>}
    }{
        \IfNoValueTF{#3}{\rep<#2||#4>}{\rep<#2|#3\rep|#4>}
    }
}

\NewDocumentCommand{\cg}{m m m}{\rep<#1; #2||#3>}

\NewDocumentCommand{\field}{o m e{_} e{^} o e{_} e{^}}{
\IfValueTF{#5}{\overline{
  #2\IfValueT{#3}{_#3}\IfValueT{#4}{^{\otimes #4}} %
  \otimes
  #5\IfValueT{#6}{_#6}\IfValueT{#7}{^{\otimes #7}} %
  \IfValueT{#1}{;#1}
}}{
  \IfValueTF{#4}{\overline{
     #2\IfValueT{#3}{_#3}\IfValueT{#4}{^{\otimes #4}}
     \IfValueT{#1}{;#1}
  }}
  {#2\IfValueT{#3}{_#3}}
}
}

\NewDocumentCommand{\frho}{o e{_} e{^}}{
\field[#1]{\rho}_{#2}^{#3}
}

\newcommand{\br}{\mbf{r}}
\newcommand{\bx}{\mbf{x}}
\newcommand{\bxhat}{\hat{\mbf{x}}}

\newcommand{\e}{a}  %

\NewDocumentCommand{\ex}{e_}{
\IfValueTF{#1}{\e_{#1}\bx_{#1}}{\e\bx}
}  %

\NewDocumentCommand{\lm}{e_}{
\IfValueTF{#1}{l_{#1}m_{#1}}{lm}
}
\NewDocumentCommand{\nlm}{e_}{
\IfValueTF{#1}{n_{#1}\lm_{#1}}{n\lm}
}
\NewDocumentCommand{\enlm}{e_}{
\IfValueTF{#1}{\e_{#1}\nlm_{#1}}{\e\nlm}
}
\NewDocumentCommand{\en}{e_}{
\IfValueTF{#1}{\e_{#1}n_{#1}}{\e n}
}
\NewDocumentCommand{\nlk}{e_}{
\IfValueTF{#1}{n_{#1}l_{#1}k_{#1}}{nlk}
}
\NewDocumentCommand{\enlk}{e_}{
\IfValueTF{#1}{\e_{#1}\nlk_{#1}}{\e\nlk}
}
\NewDocumentCommand{\enl}{e_}{
\IfValueTF{#1}{\en_{#1}l_#1}{\en l}
}

\NewDocumentCommand{\nnl}{s}{
\IfBooleanTF{#1}{n_1 n_2 l}{n_1; n_2; l}
}
\NewDocumentCommand{\ennl}{s}{
\IfBooleanTF{#1}{\en_1 \en_2 l}{\en_1; \en_2; l}
}

\NewDocumentCommand{\gslm}{s}{
\IfBooleanTF{#1}{\sigma\lambda\mu}{\sigma;\lambda\mu}
}

\newcommand{\Rhat}{\hat{R}}
\newcommand{\That}{\hat{t}}

\newcommand{\nmax}{n_\text{max}}
\newcommand{\lmax}{l_\text{max}}

\newcommand{\rcut}[0]{{r_\text{cut}} }

\fi %

\ifdefined\COSMOMODELS
\else
\newcommand{\COSMOMODELS}{}
\usepackage{bm,upgreek}

\newcommand{\krn}[0]{\operatorname{k}}

\newcommand{\bK}{\mbf{K}}

\fi %

\newcommand{\Hhat}{\hat{H}}
\newcommand{\ncent}{N}
\NewDocumentCommand\te{s}{\tilde{\e}\IfBooleanTF{#1}{'}{}}
\NewDocumentCommand\tn{s}{\tilde{n}\IfBooleanTF{#1}{'}{}}
\NewDocumentCommand\tl{s}{\tilde{l}\IfBooleanTF{#1}{'}{}}
\NewDocumentCommand\tm{s}{\tilde{m}\IfBooleanTF{#1}{'}{}}
\NewDocumentCommand\tlm{s}{\IfBooleanTF{#1}{\tl*\tm*}{\tl\tm}}
\NewDocumentCommand\tnlm{s}{\IfBooleanTF{#1}{\tnl*\tm*}{\tnl\tm}}
\NewDocumentCommand\tnl{s}{\IfBooleanTF{#1}{\tn*\tl*}{\tn\tl}}
\NewDocumentCommand\tnnlammu{}{\tnl;\tnl*;\lambda\mu}
\newcommand{\qblock}{Q}

\newcommand{\todorev}[1]{{}}
\makeatletter
\AtBeginDocument{\let\LS@rot\@undefined}
\makeatother
\begin{document}

\newcommand{\rev}[1]{#1}

\setcitestyle{super}

\title{Equivariant representations for molecular \\ Hamiltonians and $\ncent$-center atomic-scale properties}
\author{Jigyasa Nigam}
\affiliation{Laboratory of Computational Science and Modeling, Institute of Materials, \'Ecole Polytechnique F\'ed\'erale de Lausanne, 1015 Lausanne, Switzerland}
\affiliation{National Centre for Computational Design and Discovery of Novel Materials (MARVEL), {\'E}cole Polytechnique F{\'e}d{\'e}rale de Lausanne, 1015 Lausanne, Switzerland}

\author{Michael J.~Willatt}
\affiliation{Laboratory of Computational Science and Modeling, Institute of Materials, \'Ecole Polytechnique F\'ed\'erale de Lausanne, 1015 Lausanne, Switzerland}

\author{Michele Ceriotti}
\email{michele.ceriotti@epfl.ch}
\affiliation{Laboratory of Computational Science and Modeling, Institute of Materials, \'Ecole Polytechnique F\'ed\'erale de Lausanne, 1015 Lausanne, Switzerland}
\affiliation{National Centre for Computational Design and Discovery of Novel Materials (MARVEL), {\'E}cole Polytechnique F{\'e}d{\'e}rale de Lausanne, 1015 Lausanne, Switzerland}

\onecolumngrid
\newcommand{\mc}[1]{{\color{blue}{ #1}}}
\newcommand{\jn}[1]{{\color{red}{ #1}}}

\begin{abstract}

Symmetry considerations are at the core of the major frameworks used to provide an effective mathematical representation of atomic configurations that is then used in machine-learning models to predict the properties associated with each structure.
In most cases, the models rely on a description of atom-centered environments, and are suitable to learn atomic properties, or global observables that can be decomposed into atomic contributions.
Many quantities that are relevant for quantum mechanical calculations, however -- most notably the single-particle Hamiltonian matrix when written in an atomic-orbital basis -- are \emph{not} associated with a single center, but with two (or more) atoms in the structure.
We discuss a family of structural descriptors that generalize the very successful atom-centered density correlation features to the $\ncent$-centers case, and show in particular how this construction can be applied to efficiently learn the matrix elements of the (effective) single-particle Hamiltonian written in an atom-centered orbital basis.
These $\ncent$-centers features are fully equivariant -- not only in terms of translations and rotations, but also in terms of permutations of the indices associated with the atoms -- and are suitable to construct symmetry-adapted machine-learning models of new classes of properties of molecules and materials.
\end{abstract}
\twocolumngrid

\maketitle

\section{Introduction}

Most of the successful and widely used machine learning schemes that have been applied, over the past decade, to chemistry and materials aim at learning molecular energies or interatomic potentials\cite{behl-parr07prl,bart+10prl,rupp+12prl,deri+21cr,behl21cr}. %
As a consequence, the \emph{representations} that map atomic configurations into vectors of descriptors or features,\cite{SOAP,SNAP,shap16mms,ferre2017graph-kernels,eick+17nips,glie+18prb,fabe+18jcp,will+19jcp,drau20prb} to be used as inputs of the models, reflect some of the fundamental properties of the interatomic potential, such as the invariance to permutation between identical atoms, rigid rotation and inversion of the molecular structure, as well as the notions of locality and nearsightedness\cite{prod-kohn05pnas} of many components of the interatomic energy. 
This latter, in particular, suggested to use atom-centered features that describe the arrangement of neighbors around a tagged atom. The number of neighbors considered simultaneously enumerates a hierarchy of $\nu$-order correlation representations, with increasing complexity and descriptive power. 
Notably, atom-centered representations have been used not only to build models of properties associated with an individual atomic center $i$, such as NMR chemical shieldings~\cite{gran-harr96book,pick-maur01prb,paru+18ncomm}, but also to express global, extensive properties such as the molecular energy as a sum of atom-centered contributions.

More recently, symmetry-invariant models have been increasingly generalized to be imbued with \emph{equivariant} behavior with respect to rotations and inversion,\cite{glie+17prb,gris+18prl,ande+19nips,unke-meuw19jctc}, addressing the need to construct data-driven models for atomic properties that have a structure more complicated than that of a scalar, such as dipole moments,\cite{veit+20jcp,zhan+20prb} polarizabilties,\cite{wilk+19pnas} and fields such as the charge\cite{gris+19acscs} and on-top\cite{fabr+20jcp} densities. 
In the vast majority of cases, these equivariant representations are still used together with atom-centered frameworks, that allow for superior transferability between systems of different size -- the main limitation being connected with the finite range of the environment they can describe\cite{yue+21jcp}, unless they are combined with features specifically designed to capture the multi-scale nature of interactions\cite{gris+21cs}.

While atom-centered descriptions are the most common in the construction of interatomic potentials, there are also examples in which the energy has been expanded as a sum of \emph{pair} energies\cite{jose+12jcp}.
Perhaps more importantly, there are several properties that are intrinsically associated with multiple atomic centers. 
J-couplings in NMR\cite{joyc+07jcp}, that describe the magnetic interaction between nuclear spins, are a typical example. 
Another example, that we will focus on in this work, involves the matrix elements of \rev{a one-particle, effective} electronic Hamiltonian ($\Hhat$) when written in an atomic orbital basis. 
A classical example is that of tight-binding models\cite{sutt+88jphc}, in which the matrix elements of a minimal-atomic-basis Hamiltonian are parameterized in terms of interatomic distances and angles. 
Constructing data-driven models that match more closely explicit electronic structure calculations is an appealing approach to obtain improved semiempirical methods, and to access the many observables, such as optical excitations\cite{west-maur21cs}, that can be approximated based on the eigenvalues of an effective single-particle Hamiltonian. Furthermore, a machine-learned $\Hhat$ could also be used as inputs of the emerging family of ML models that predict molecular properties using matrix elements computed by explicit electronic-structure calculations\cite{welborn2018transferability,qiao2020orbnet} or the corresponding eigenvalues\cite{fabrizio2021}.

Existing machine learning approaches that attempt to predict the molecular Hamiltonian do so in terms of an ad-hoc modification of the atom-centered features \cite{hegde2017machine} or by devising pair features\cite{schu+19nc}, that however do not explicitly include the rotational symmetries and instead rely on data augmentation to incorporate them in the model. An interesting approach, recently proposed by Westermayr and Maurer,\cite{west-maur21cs} predicts an \emph{invariant} pseudo-Hamiltonian matrix that has the correct eigenvalues, even though it does not need to bear any resemblance to the actual atomic orbital Hamiltonian. \rev{A very recent proposal of an SE(3) equivariant neural network model of the Hamiltonian has demonstrated dramatic improvements over non-symmetric models\cite{unke2021se3equivariant}.}

To put the problem of learning these kinds of properties on a more solid mathematical footing, we introduce a symmetrized $\ncent$-center representation that provides a natural, fully equivariant framework to learn properties that are associated with $\ncent$ atoms. We discuss its construction and its relationship to 1-center density correlation representations, and then demonstrate its use for the particular case of the two-center Hamiltonian.  
We show that the combination of atomic index and geometric equivariance incorporates naturally the symmetry considerations behind molecular orbital theory. As a result of the symmetry constraints, a small number of reference configurations suffice to achieve robust, accurate predictions of $\Hhat$, that are competitive with state-of-the-art, deep-learning models despite using only linear or kernel regression.

\begin{figure}
    \centering
    \includegraphics[width=0.8\linewidth]{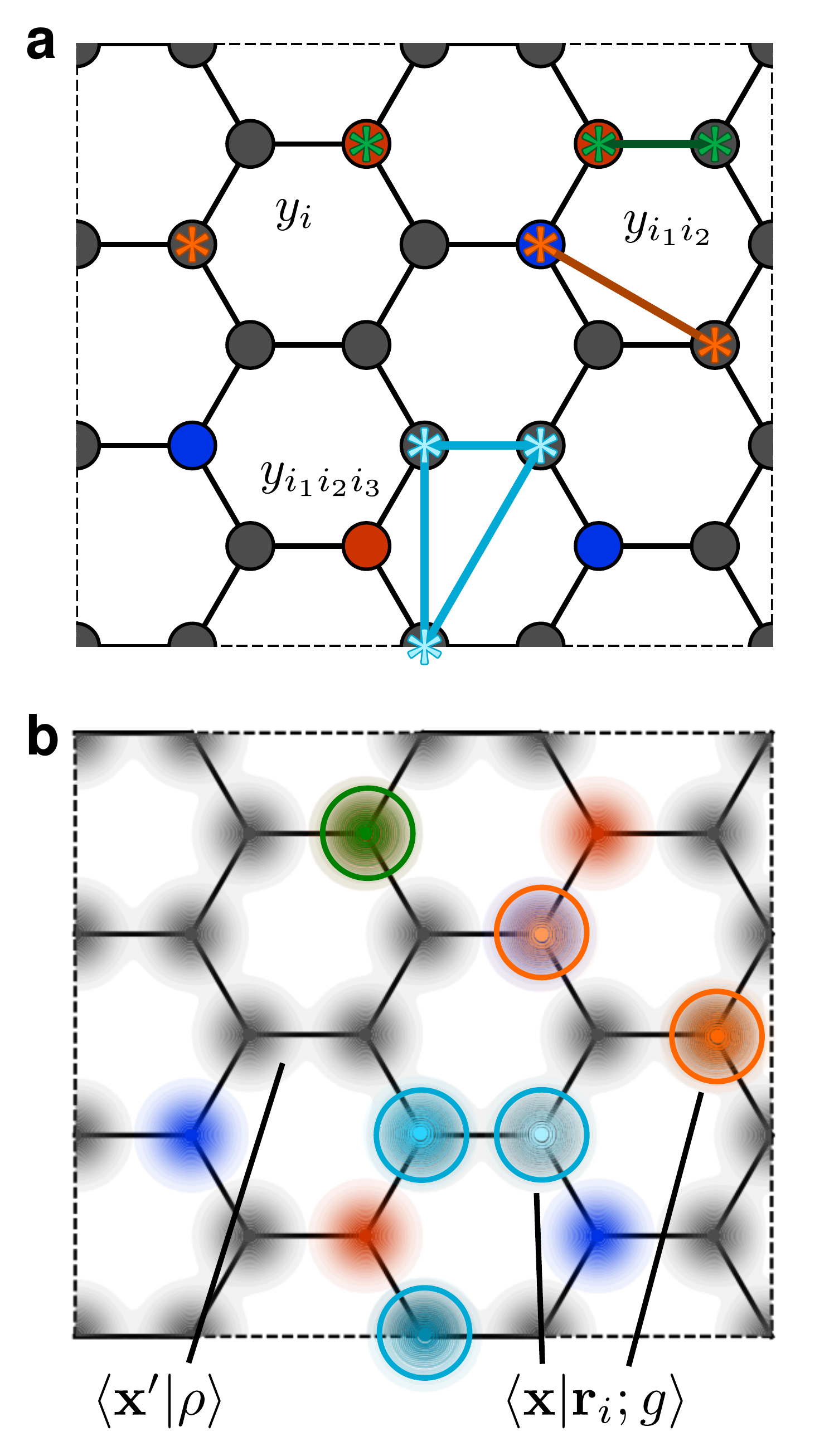}
    \caption{ \rev{ (a) A schematic depiction of the definition of $N$-centers quantities associated with an atomic configuration. The tagged atoms in each cluster are associated with 1,2, and 3-centers properties. (b) A representation of the same structure in terms of atom density fields. The structure is described as a decorated atom density, with different chemical elements associated with a different type of density: this description is invariant to permutations of atoms of the same type. Each $N$-center group can be identified by adding one separate atom-centered Gaussian to each atom in the cluster, which are represented as colored Gaussians matching the clusters in the top panel.  } }
    \label{fig:n-centers}
\end{figure}

\section{$\ncent$-centers features}

We address the problem of building a representation for atomistic properties that depend on the indices of several atoms, $y_{i_1 i_2\ldots}$ of a structure $A$ (Figure~\ref{fig:n-centers}a).
Atomic properties may be written as $y_i$, and properties associated with pairs of atoms as $y_{i_1i_2}$.
Models built to predict these properties should be equivariant with respect to changes in the atom labels:
if we relabel the indices 1 and 2, respectively, as 7 and 9, equivariance requires that $y_1\rightarrow y_7$, $y_2\rightarrow y_9$, $y_{12}\rightarrow y_{79}$, $y_{221}\rightarrow y_{997}$, etc.

We may write a representation that aims to describe the structural features that determine the value of $y_{i_1i_2\ldots}(A) $ as $\rep<q||A_{i_1i_2\ldots}>$. 
We use the bra-ket notation that has been introduced in Refs.~\citenum{will+18pccp,will+19jcp}; the interested reader will find a thorough discussion in Section IV.A of Ref.~\citenum{musi+21cr}.
In short, the ket indicates the entity being represented (the $N$-centers tuple $A_{i_1i_2\ldots}$), supplemented with other indices that describe the nature and symmetry of the descriptors. 
The bra, instead, contains the (discrete or continuous) indices that enumerate the components of the feature vector associated with this entity. 
A linear model to predict $y_{i_1i_2\ldots}$ can then be written as
\begin{equation}
\tilde{y}_{i_1i_2\ldots}(A) = \sum_q \rep<y||q> \rep<q||A_{i_1i_2\ldots}>.
\end{equation}
It is clear that -- in order for the prediction $\tilde{y}$ to be equivariant with respect to permutation of indices -- the features associated with $A_{i_1i_2\ldots}$ must also be equivariant to changes in the atom labeling. 
In fact, the atom-centered features that underlie the vast majority of modern machine-learning interatomic potentials \emph{are} equivariant with respect to the atom labelling: if we reverse the storage order of the atoms in a structure, the storage order of the various atomic contributions will also  be reversed. 
Summing over the contributions to yield the \emph{total} energy leads to a permutation-invariant model, and the equivariance of the underlying contributions is not usually given much emphasis. 

\subsection{Label-equivariant structural representation}
\begin{figure}[tbp]
\includegraphics[width=1.0\linewidth]{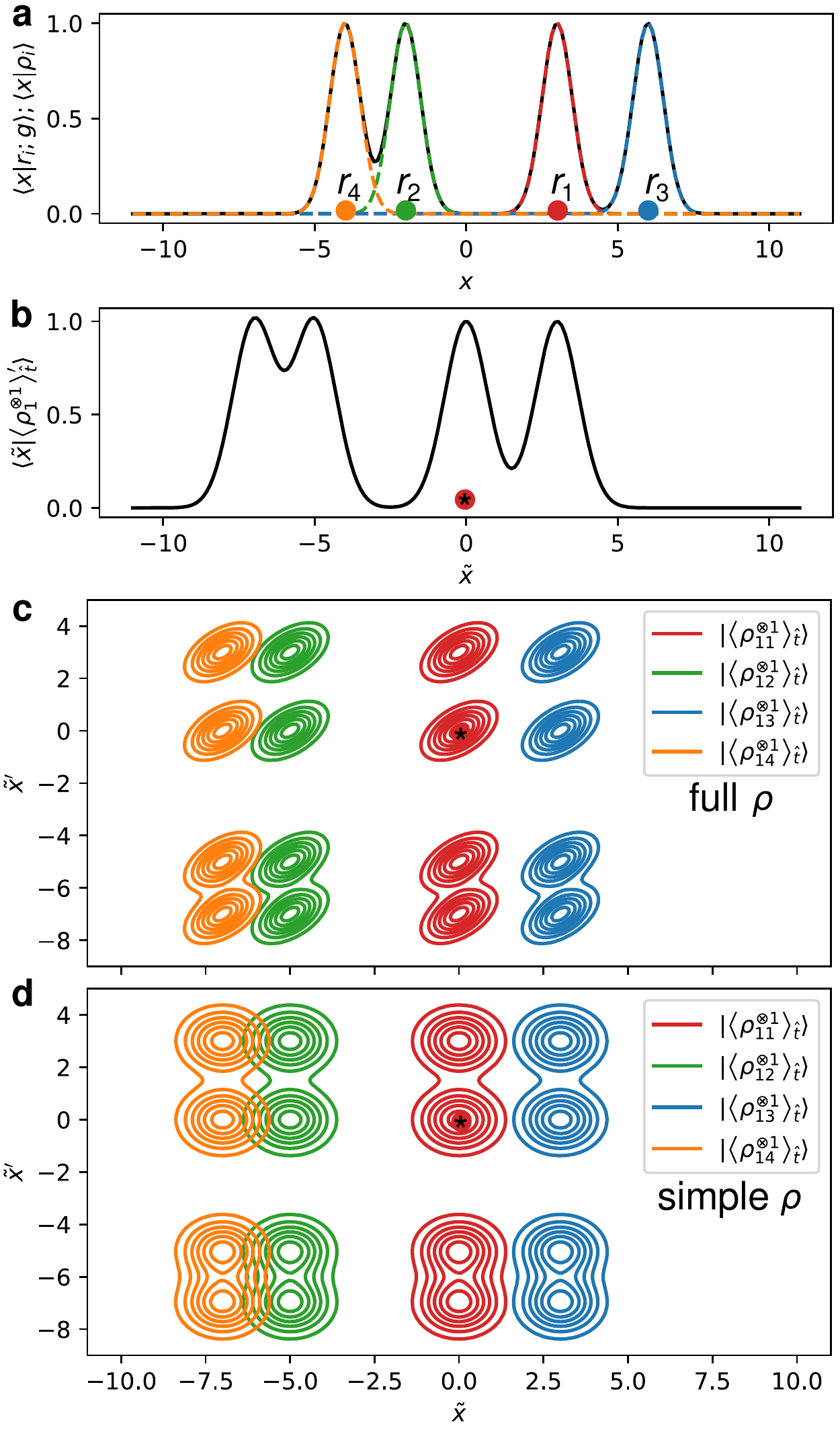}
\caption{
(a) A 1D group point cloud, represented together with the density contributions and the total density $\rep<x||\rho>$. 
(b) The translationally-symmetrized density $\rep<x||\rho_i>$ centered on the atom $i=1$. 
(c) Translationally-symmetrized pair features $\rep<\tilde{x};\tilde{x}'|| {\left<\rho_{ii'}^{\otimes 1}\right>_{\hat{t},\text{full}}} >$, for $i=1$, and with different $i'$ represented with different colors. (d) As in (c), for the simplified form in Eq.~\eqref{eq:rho-trans}.
\label{fig:rho-trans}
}
\end{figure}

We can write descriptors that fulfill atom-indices equivariant properties by ``tagging'' the selected atoms with 
an atom-centered function, e.g. a Gaussian $\rep<\bx||\br_i; g>\equiv g(\bx-\br_i)$.
Each center can be associated with a separate function, so that the descriptor is indexed by $3N$ spatial coordinates, and the relative position of the atoms is encoded in the position of the Gaussian peaks.
In order to express the relationship between the tagged centers and the rest of the structure, we can combine these functions with \rev{$\nu$ copies of} the overall atom density $\rep<\bx||\rho>=\sum_j \rep<\bx||\br_j; g>$,  that are invariant to permutations of the atoms labels
\begin{multline}
\rep<\bx_1; \bx_2; \ldots \bx_n;\bx'_1; \bx'_2 ;\ldots \bx'_\nu ||\rho_{i_1i_2\ldots i_\ncent}^{\otimes \nu}>\equiv\\ \rep<\bx_1||\br_{i_1}; g> \rep<\bx_2||\br_{i_2}; g> \ldots \rep<\bx_n||\br_{i_\ncent}; g> \times \\
 \sum_{j_1} \rep<\bx'_1||\br_{j_1}; g> 
\sum_{j_2} \rep<\bx'_2||\br_{j_2}; g> \ldots\sum_{j_\nu}\rep<\bx'_\nu||\br_{j_\nu}; g>.
\label{eq:rho-equi}
\end{multline}
The tensor products of $\rep<\bx||\rho>$ describes the structure of the environment ($\nu=1$ provides a full description at this level, but $\nu>1$ will become necessary when symmetrizing the representation), while the Gaussian centered on the $i$-atoms identify the atom tuple for which $y_{i_1i_2\ldots}$ is to be predicted (Figure~\ref{fig:n-centers}b).
Figure~\ref{fig:rho-trans}a shows another example of atom-centered Gaussians and the total atom density for a 1D point cloud.

\subsection{Translational symmetry}

Averaging Eq.~\eqref{eq:rho-equi} over the translation group leads to a rather daunting expression: the convolution of $(\ncent+\nu)$ Gaussian densities generates a product of $(\ncent+\nu)(\ncent+\nu-1)/2$ Gaussians relative to differences of the arguments of pairs of the densities (to be precise, with a variance that is increased by a factor of $(\ncent+\nu)$ relative to the initial Gaussians), 
\begin{multline}
\int \D{\That} \rep<\bx_1; \bx_2; \ldots \bx_n;\bx'_1; \bx'_2 ;\ldots \bx'_\nu |\That \rep|\rho_{i_1i_2\ldots i_\ncent}^{\otimes \nu}>\equiv\\[-1ex]
\sum_{j_1j_2\ldots }\prod_{\substack{
\alpha\in \{i_1\ldots j_1\ldots\}\\
\beta\in \{i_1\ldots j_1\ldots\}, \beta>\alpha
}} 
\rep<\bx_\beta -\bx_{\alpha}||\br_\beta -\br_\alpha ; g>, 
\end{multline}
where it is implied that $\bx_{i_\alpha} \equiv \bx_\alpha$ and $\bx_{j_\alpha} \equiv \bx_\alpha'$. 
\newcommand{\tbx}{\tilde{\bx}}
Even though the result contains a large number of Gaussian factors, it only functionally depends on $(\ncent+\nu-1)$ differences between the feature indices, $(\bx_\beta - \bx_\alpha)$ (see Fig.~\ref{fig:rho-trans}b for the simplest case where only the atom density $\rep<x||\rho>$ is symmetrized over translations). We can pick the $i\equiv i_1$ atom as reference, define $\tbx_i=\bx_i -\bx_1$ and write 
\begin{multline}
\rep<\tbx_2; \ldots \tbx_n; \tbx'_1; \tbx'_2 ;\ldots \tbx'_\nu||{\langle\rho_{i i_2\ldots i_\ncent}^{\otimes \nu}\rangle_{\That, \text{full}}} >\equiv\\
\prod_{\beta\in\{i_2\ldots i_\ncent\}}\!\!\!\!\!
\rep<\tbx_\beta||\br_{\beta i}; g>
\sum_{j_1\ldots j_\nu}\prod_{\beta\in \{j_1\ldots j_\nu\}}
\!\!\!\!\!\rep<\tbx'_\beta||\br_{\beta i}; g>
\\[-1ex]
\times\prod_{\substack{
\alpha\in \{i_2\ldots j_1\ldots\}\\
\beta\in \{i_2\ldots j_1\ldots\}, \beta>\alpha
}}\!\!\!\!\!\!\!\!\!\!
\rep<\tbx_\beta -\tbx_\alpha||\br_{\beta i} -\br_{\alpha i} ; g>.  \label{eq:rho-trans-monster}
\end{multline}
The product on the third line contains largely redundant information (as it couples distance vectors that are already in the argument of Gaussians centered on the selected $i$ atom), and is computationally problematic, as it couples the densities associated with different $j$ indices preventing the application of the density trick that can be used to evaluate these features efficiently.\cite{musi+21cr} 
Thus, we just drop it altogether, reorder the indices, and define the atom-centered, translation-invariant, permutation-equivariant features as 
\begin{multline}
\rep<\bx_2; \ldots \bx_n; \bx'_1; \bx'_2 ;\ldots \bx'_\nu||{\langle\rho_{i i_2\ldots i_\ncent}^{\otimes \nu}\rangle_{\That}} >\equiv\\
\prod_{\alpha=2}^\ncent
\rep<\bx_\alpha||\br_{i_\alpha i}; g> 
\prod_{\beta=1}^\nu
\rep<\bx'_\beta||\rho_i> \label{eq:rho-trans}
\end{multline}
where we introduced the $i$-centered density $\rep|A; \rho_i> \equiv \sum_{j\in A} \rep|\br_{ji}; g> $.
Figure~\ref{fig:rho-trans}(c-d) demonstrates, in a simple case, how Eq.~\eqref{eq:rho-trans-monster} and~\eqref{eq:rho-trans} contain similar amounts of structural information.
\rev{As we shall see, this simplified form of the translationally-symmetrized correlations can be directly related to established atom-centered density correlation features, which further motivates discarding the redundant Gaussian terms in Eq.~\eqref{eq:rho-trans-monster}.}

\begin{figure}[tbp]
\includegraphics[width=1.0\linewidth]{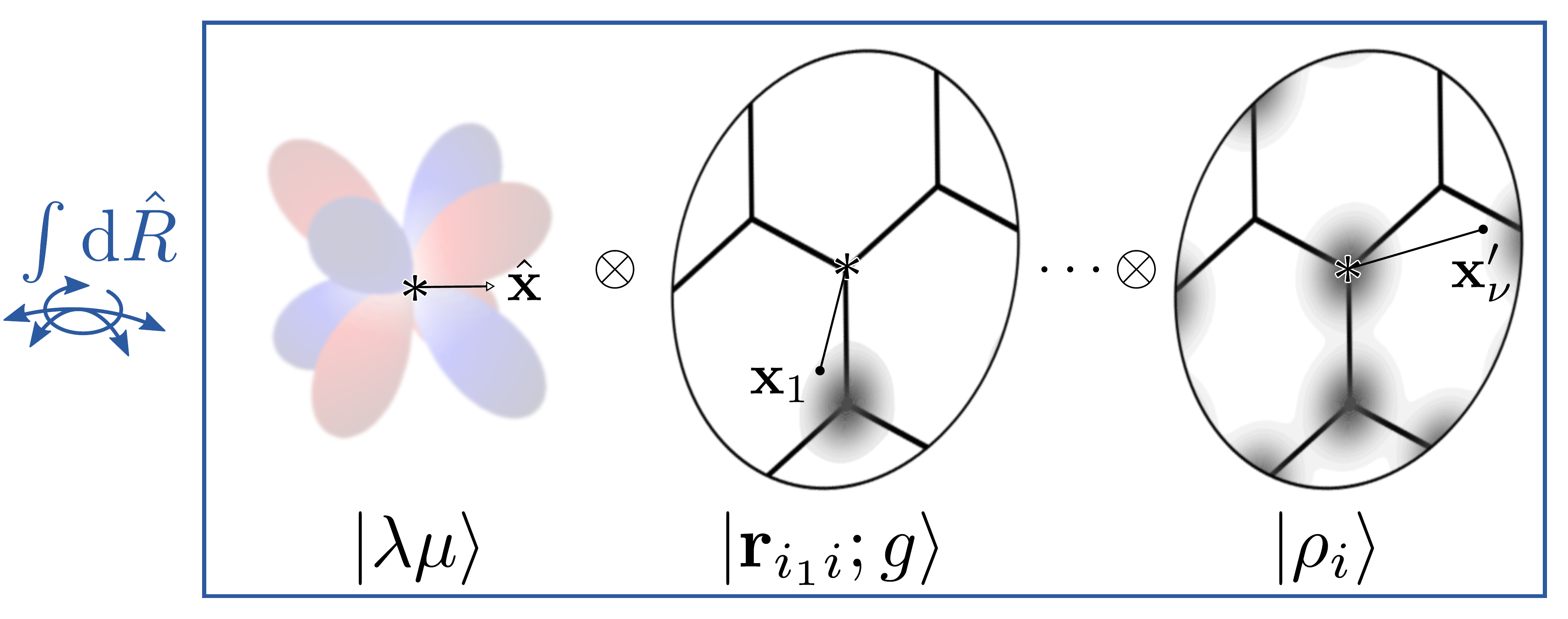}
\caption{
A schematic representation of the construction of $O(3)$ equivariant pair features, as the symmetry average of a tensor product of pair densities, a neighbor density and a set of spherical harmonics.
\label{fig:rho-lm-pair}
}
\end{figure}

\subsection{Rotational symmetry}

Having selected one reference $i$-atom makes it simple to relate Eq.~\eqref{eq:rho-trans} to the $O(3)$ equivariant features that can be obtained by symmetrizing the $\nu$-point neighbor density correlations\cite{will+19jcp}. 
For $\ncent=1$, this symmetry average amounts to 
\begin{multline}
\!\!\!\rep|\frho[\gslm]_i^\nu > \equiv
\int_{O(3)} \!\!\!\!\!\! \D{\Rhat} \prod_{\beta=1}^\nu
\Rhat\rep|\rho_i> \otimes \Rhat \rep|\sigma> \otimes \Rhat \rep|\lambda\mu>
\end{multline}
where we use $\D{\Rhat}$ to indicate averaging over proper and improper (without and with inversion) rotations, we use a compact notation that does not indicate explicitly the basis, and  $\rep|\sigma;\lambda\mu>$ tracks the transformation behavior with respect to the symmetry operation.\cite{niga+20jcp} 
Including also the Gaussians that tag further centers is formally analogous:
\begin{multline}
\!\!\!\rep|\frho[\gslm]_{{ii_2\ldots i_\ncent}}^\nu > \equiv
\int_{O(3)} \!\!\!\!\!\! \D{\Rhat} 
\prod_{\alpha=2}^\ncent
\Rhat\rep|\br_{i_\alpha i}; g> 
\prod_{\beta=1}^\nu
\Rhat\rep|\rho_i>  \\\otimes \Rhat \rep|\sigma> \otimes \Rhat \rep|\lambda\mu>,
\end{multline}
providing the general, abstract expression for the $\ncent$-centers, $\nu$-neighbors symmetrized density correlation equivariants, that is represented schematically in Fig.~\ref{fig:rho-lm-pair}.
With these definitions, one recognizes a connection between different $N$-center features, 
\begin{equation}
\!\!\!
\sum_{i_\ncent}
\rep|\frho[\gslm]_{{ii_2\ldots i_\ncent}}^\nu > 
=\rep|\frho[\gslm]_{{ii_2\ldots i_{\ncent-1}}}^{\nu+1} >.
\end{equation}
By summing over all of the centers in a structure, one eliminates the dependence of $\ncent$-center features on one of the $i$ indices, and converts them into a $(\ncent-1)$-center representation of higher body order. 
In other terms, one could take $N$-center, $\nu=0$ equivariants as the starting point of the construction, and make them invariant with respect to atom index permutations by summing over all the $i$ indices -- effectively building permutation-invariant representations by summing over $\nu$-center clusters (as it is done for instance in the calculation of atom-centered symmetry functions\cite{behl11jcp}) rather than by symmetrized products of $\rep|\frho_i>$.\cite{musi+21cr}

To compute these descriptors in practice, it suffices to expand the neighbor densities in radial functions $\rep<x||nl>$ and spherical harmonics $\rep<\bxhat||lm>$ 
\begin{equation}
\rep<nlm||\br_{ji} ; g> \equiv 
\int \D{\bx} \rep<nl||x> \rep<lm||\bxhat> \rep<\bx||\br_{ji} ; g>,
\end{equation}
and apply expressions that are entirely analogous to those for the $(\ncent+\nu)$-neighbors density correlation features.\cite{will+19jcp,niga+20jcp,musi+21cr}
In most implementations, including the one we use here, one usually adopts real spherical harmonics, which requires using a consistent definition of the angular momentum sum rules.\cite{gosc+21jcp} 
The expressions we write in the text assume real-valued coefficients.
Considering invariant pair features $\rep|\frho_{{ii'}}^\nu>$ one gets the two-centers, zero-neighbors term
\begin{equation}
\rep<n||\frho_{{ii'}}^0> \equiv \rep<n 00 ||\br_{i'i}; g>,
\end{equation}
the two-centers, one-neighbor features
\begin{equation}
\rep<n_1 n_2 l||\frho_{{ii'}}^1> \equiv  \sum_{m} \frac{1}{\sqrt{2l+1}} \rep<n_1 l m ||\br_{i'i}; g> \rep<n_2l m ||\rho_i>,
\end{equation}
or the 3-centers, zero-neighbor features
\begin{equation}
\rep<n_1 n_2 l||\frho_{{ii_2 i_3}}^0> \equiv  \sum_{m} \frac{1}{\sqrt{2l+1}} \rep<n_1 l m ||\br_{i_2i}; g> \rep<n_2 l m ||\br_{i_3i}; g>,
\end{equation}
and so on and so forth. 

Analogous constructions can be applied to other atom-centered frameworks, such as moment tensor potentials\cite{shap16mms}, or the atomic cluster expansion\cite{drau20prb} -- the latter having been used as the basis for a similar, independent effort\cite{zhang+21arxiv}. 
Given the fundamental equivalence with the density-correlation frameworks, it may also be possible to extend atom-centered symmetry functions\cite{behl11jcp} or the FCHL features\cite{fabe+18jcp} in the same direction.
Higher-body order equivariant features can be easily obtained within the $N$-body iterative contraction (NICE) framework\cite{niga+20jcp}, applying an iteration of the form
\begin{multline}
\rep<\cdots; nlk||\frho[\sigma;\lambda\mu]_{{i\cdots i_N}}^{(\nu+1)}>\\ \propto
\sum_{m h} \cg{lm}{kh}{\lambda \mu}
\rep<n l m||\rho_i> \\[-1ex]
\times\rep<\cdots||\frho[(\sigma (-1)^{l+k+\lambda});k h]_{{i\cdots i_N}}^{\nu}> 
\label{eq:nice-body}
\end{multline}
to increase the order of the neighbor-density description, and
\begin{multline}
\rep<\cdots; nlk||\frho[\sigma;\lambda\mu]_{{i\cdots i_Ni_{N+1}}}^{\nu}>\\ \propto
\sum_{m h} \cg{lm}{kh}{\lambda \mu}
\rep<n l m||\br_{i_{N+1} i}; g> \\[-1ex]
\times \rep<\cdots||\frho[(\sigma (-1)^{l+k+\lambda});k h]_{{i\cdots i_N}}^{\nu}> 
\label{eq:nice-center}
\end{multline}
to include an additional center in the representation. In the example we show here, we restrict ourselves to features up to $(N=2, \nu=2)$, but the extension to arbitrary $N$ and $\nu$ poses no conceptual challenge.

\subsection{Index permutation symmetry}

Expressing the translationally-symmetrized $\ncent$-center features in a compact, non-redundant basis, and selecting one $i$-atom as the origin for their construction 
(that is, discarding the cumbersome term in \eqref{eq:rho-trans-monster}) simplifies greatly their evaluation and manipulation. 
However, it obscures the symmetry of the features with respect to permutations of the $\ncent$-center indices, which can be problematic when one wants to learn e.g. a 2-center quantity that is symmetric with respect to a swap of the $i$ indices. 
To address this further symmetry, we define permutation-symmetrized $\ncent$-center features
\begin{equation}
\rep|\frho[\pi]_{{i_1i_2\ldots i_\ncent}}^{\nu} > \equiv
\sum_{\mbf{p}\in\operatorname{perm}(\ncent)}
\operatorname{sign}_\pi(\mbf{p}) \rep|\frho_{{i_{p_1}i_{p_2}\ldots i_{p_\ncent}}}^{\nu} >. \label{eq:rho-perm}
\end{equation}
For example, the symmetric and antisymmetric pair features can be easily obtained as
\begin{equation}
\rep|\frho[\alpha;\pm]_{{ii'}}^{\nu}> = 
\rep|\frho[\alpha]_{{ii'}}^{\nu}> \pm \rep|\frho[\alpha]_{{i'i}}^{\nu}>,
\label{eq:rho-perm-pair}
\end{equation}
where $\alpha$ stands for any additional symmetry index, such as parity $\sigma$ or rotational indices $\lambda\mu$.

\section{Learning atomic-orbital matrices}

As a practical application of the $\ncent$-center representations, let us consider the case of modeling the matrix elements of \rev{an effective one-electron} electronic Hamiltonian, written in a basis of atom-centered orbitals 
\begin{equation}
\rep<\hat{H}; \e_i\tnlm; \e_{i'}\tnlm*||A_{ii'}> \equiv\bra*{i\tnlm}\Hhat\ket*{i'\tnlm*}.
\label{eq:hamiltonian-decoupled}
\end{equation}
We use a tilde decoration to distinguish the orbital indices from the indices used to enumerate pair features, and we recast the usual matrix-element notation from quantum mechanics into one that is more suitable to enumerate the learning targets in a ML exercise. 
The \rev{one-electron} Hamiltonian is just one of the many matrices describing pair terms in electronic-structure theory, and other entities that could be manipulated in similar ways include the density matrix, and the response of the Hamiltonian to external perturbations. 
The bra on the left-hand side enumerates the entries in the Hamiltonian matrix, which can be broken down into blocks, each of which is associated with a distinct model that should be separately trained  to be able to reconstruct the full matrix. 
For each \emph{type} of blocks (atomic species, radial channels, ...) we are going to build models that take as inputs the  features associated with a pair $A_{ii'}$ and predict the corresponding section of the Hamiltonian. 
To give a concrete example of the architecture we are considering, a linear model to predict the matrix element between the $2s$ and $1s$ orbitals on a O and H atom can  be written as
\begin{equation}
\rep<\hat{H}; \ce{O}200; \ce{H}100 ||A_{ii'}> \approx
\sum_q \rep<\hat{H}; \ce{O}2s; \ce{H}1s ||q> \rep<q||A_{ii'}; 00>,
\end{equation}
where $q$ indicates the discretization of the feature vector. The regression weights are specific to the type of the block ($\ce{O}2s; \ce{H}1s$), while the features describe the two selected atoms and their environment, and have the appropriate equivariance ($\lambda=0$) for the orbitals under consideration. 
While it would be possible, and probably beneficial, to adapt the hyperparameters of the features depending on the orbital type, we will use the same set of features for all blocks with equivalent symmetry, e.g. we use exactly the same $\rep<q||A_{ii'}; 00>$ when predicting the ($\ce{O}1s; \ce{H}1s$) block type. 

\begin{figure}
    \centering
    \includegraphics[width=1.0\linewidth]{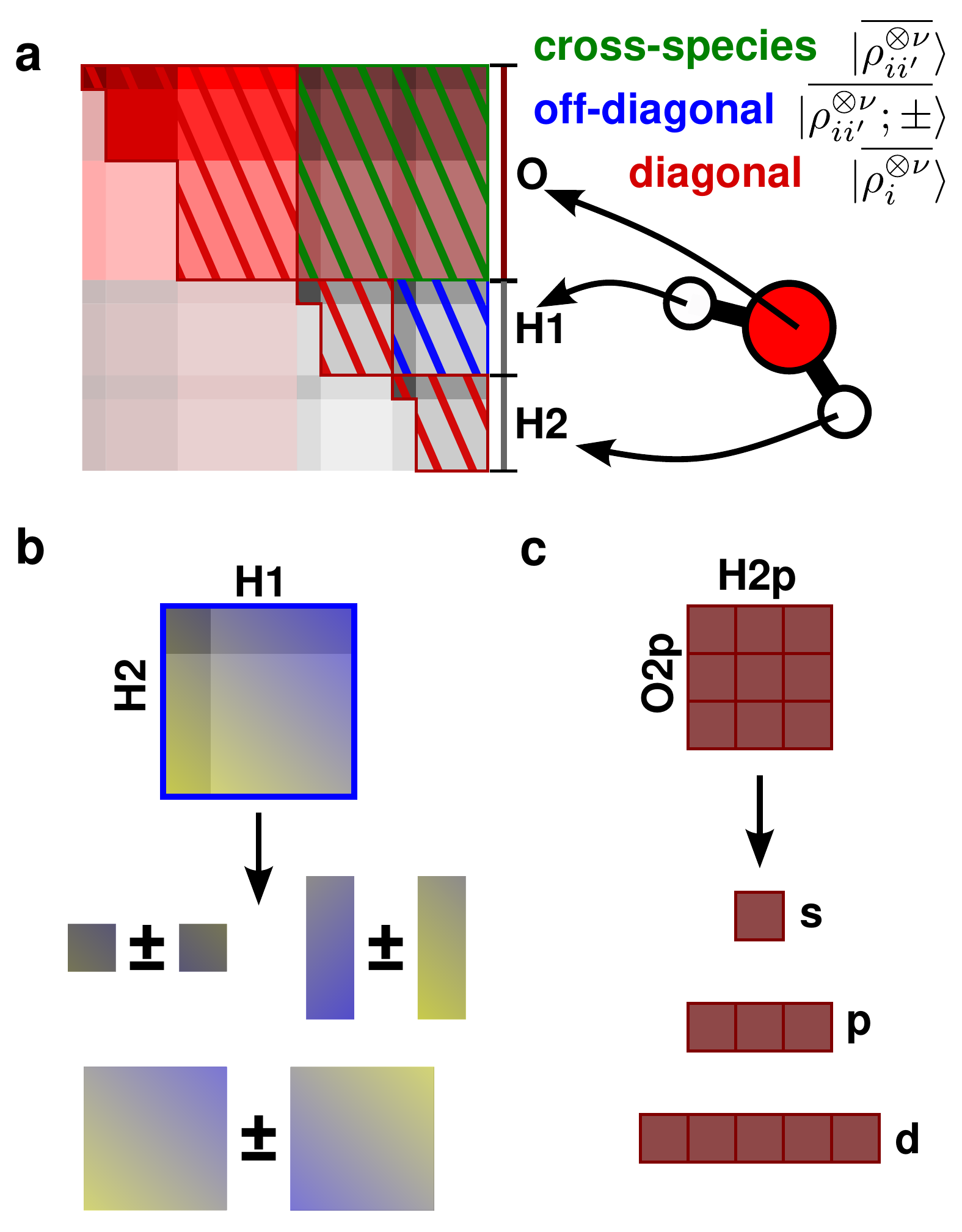}
    \caption{(a) A schematic representation of the different types of blocks that compose the Hamiltonian for a water molecule. (b) Blocks involving different atoms of the same species must be transformed into the index-permutation equivariant forms. (c) The $O(3)$ equivariance of different angular momentum blocks can be simplified by converting them into a coupled basis, separating their irreducible components.}
    \label{fig:ham-blocks}
\end{figure}

\subsection{Permutation equivariance}

The entries in the Hamiltonian are not fully symmetric, and it is advisable to convert them in an irreducible form. First, we need to consider the symmetry with respect to the swaps of atom and orbital labels. 
The simultaneous swap of atom and orbital corresponds to the Hermitian symmetry of the Hamiltonian:
\begin{equation}
\rep<\hat{H}; \e_{i}\tnlm; \e_{i'}\tnlm*||A_{ii'}> = \rep<\hat{H}; \e_{i'}\tnlm*; \e_{i}\tnlm;|| A_{i'i}>^*.
\end{equation}
In fact, we will work with Hamiltonians written using real-valued spherical harmonics, and so the matrix is real and symmetric. 
This means that only half of the Hamiltonian needs to be predicted. To obtain perfect equivariance with respect to the sorting of atomic indices, one should proceed with care in handling blocks associated with different types of atoms. 

\paragraph*{ Cross-element terms. } Consider first the case of matrix elements between atoms of different species. We can always choose to pick the atoms in a prescribed order, e.g. consider the heavier element as the first in the pair. 
Still, we need a separate model for each combination of orbitals on the two centers: there is no symmetry relationship between $(\e \tnl; \e'\tnl*)$ and $(\e \tnl*; \e'\tnl)$, even if one used the same radial basis for both elements, because e.g. the matrix element for a $1s$ orbital centered on O and a $2s$ orbital centered on H is different from the matrix element for a $2s$ orbital centered on O and a $1s$ orbital centered on H.
The element types that define the block type also induce a natural ordering of the pair -- that is, one can choose consistently the order of the indices $(i,i')$ in the pair features. 
Thus, there is no need to use features with a specified particle-exchange symmetry -- one could use either $\rep|\frho_{{ii'}}^{\nu}>$, $\rep|\frho_{{i'i}}^{\nu}>$. To incorporate the maximum amount of information, and to have a scheme that is independent on an arbitrary choice of ordering, we concatenate the two asymmetric pair features $\rep|\frho_{{ii'}}^{\nu}> \oplus \rep|\frho_{{i'i}}^{\nu}>$. 

\paragraph*{ Same-element, off-diagonal terms. } 
These terms require particular attention. Contrary to the previous case, one cannot choose a priori the order of the atoms in the pair, because they are indistinguishable. 
The matrix elements, however, are not symmetric with respect to the swap of atom indices $(i,i')$. To give a concrete example $\rep<\Hhat; \ce{O}100;\ce{O}200||A_{37}>$ has no prescribed relationship to $\rep<\Hhat; \ce{O}100;\ce{O}200||A_{73}>$, because the environment of the \ce{O} atom at $\br_7$ might be completely different from that at $\br_3$. 
When given a new structure to predict, however, we only know there are two \ce{O} centers and we want to predict a matrix element for $(\ce{O}100;\ce{O}200)$, so the weights would be the same for $A_{37}$ and $A_{73}$. 
Even though the asymmetric $\rep|\frho_{{37}}^{\nu}>$ and $\rep|\frho_{{73}}^{\nu}>$ differ, there is no reason they should -- when combined with the \emph{same set of weights} -- give meaningful predictions for the two sets of elements of $\Hhat$. In other terms, given that we cannot fix the order of the two centers to match the orbital blocks associated with them, we cannot build a model that depends on the ordering of the atoms of each species.

To address this issue, we need to modify the target so that it is equivariant with respect to a swap of the center indices. 
This can be achieved by building symmetric and antisymmetric combinations of the entries
\begin{multline}
\rep<\hat{H}; \e_{i}\tnlm; \e_{i}\tnlm*; \pm|| A_{ii'}>
\equiv\\
\rep<\hat{H}; \e_{i}\tnlm; \e_{i}\tnlm*||A_{ii'}> \pm \rep<\hat{H}; \e_{i}\tnlm; \e_{i}\tnlm* || A_{i'i}>.
\end{multline}
Given that these combinations transform in a precise way under an exchange of the particle indices, they can be suitably learned with the symmetric and antisymmetric pair features~\eqref{eq:rho-perm-pair}, $\rep|\frho[\pm]_{{ii'}}^{\nu}>$.
Furthermore, one can exploit Hermitian symmetry, and learn only models for lexicographically-ordered $(\tnl,\tnl*)$ terms, i.e. for $\tn*\ge\tn$, and (when $\tn*=\tn$) $\tl*\ge\tl$. 
Terms that correspond to $(\tn*,\tl*)=(\tn, \tl)$ are special, because swapping the atom indices corresponds to taking the transpose of the block. As a consequence, due to the Hermitian property of the Hamiltonian, some of the symmetrized terms are bound to be zero (see the SI).

\paragraph*{ Diagonal terms. } Terms that correspond to on-site blocks, $i=i'$ should not be considered pair terms at all, and are best learned using one-center features  $\rep|\frho_{{i}}^{\nu}>$. Similar to the same-species case, thanks to the Hermitian symmetry, we only need to build models for the lexicographically ordered orbital pairs.

\subsection{$O(3)$ equivariance}
\label{sub:ham-equivariance}

The Hamiltonian block $\rep<\hat{H}; \tnlm; \tnlm*||A_{ii'}>$ transforms as a product of (real) spherical harmonics,\footnote{The coupling relations we write in this section and elsewhere are formally equivalent to the usual relationships that exist for complex-valued spherical harmonics, with the understanding that the CG coefficients need to be adapted accordingly.} $\rep|\tl \tm> \otimes \rep|\tl* \tm*>$. 
We omit the indication of the nature of the elements, and the possible symmetrization with respect to atom indices permutation, because the same arguments apply to each type of blocks discussed in the previous paragraph. 
We can use well-known relationships between products and sums of spherical harmonics to convert the matrix elements into irreducible representations of $SO(3)$:
\begin{equation}
\begin{split}
\!\!\!\rep<\hat{H}; \tnnlammu||A_{ii'}> = &
\sum_{\tm\tm*} \!\rep<\hat{H}; \tnlm; \tnlm*||A_{ii'}> \\ &\times \cg{\tl\tm}{\tl*\tm*}{\lambda\mu}
\end{split}\label{eq:hamiltonian-coupled}
\end{equation}
and back
\begin{equation}
\begin{split}
\rep<\hat{H}; \tnlm; \tnlm*||A_{ii'}> = 
\sum_{\lambda\mu} 
\rep<\hat{H}; \tnnlammu||A_{ii'}>\\[-2ex]
\times 
\cg{\tl\tm}{\tl*\tm*}{\lambda\mu}. 
\end{split}
\end{equation}
This is preferable to building symmetry-adapted models for the uncoupled basis (which would be possible with minor modifications of the equivariant construction) because we can re-use the same framework that is routinely applied to the learning of tensorial properties\cite{gris+18prl} -- i.e. the symmetry-adapted equivariant features, extended here to multi-center $\rep|\frho[\lambda\mu ]_{{i\ldots i_N}}^{\nu}>$ --  and because the learning of large blocks of geometrically covariant terms is broken down into smaller terms that correspond to irreducible representations of $O(3)$.
For the off-diagonal, same species blocks, the coupled-basis form of the Hamiltonian matrix elements can then be symmetrized with respect to atom-index exchange, yielding quantities of the form $\rep<\hat{H}; \tnnlammu; \pm||A_{ii'}> $ that can be learned with the corresponding symmetric and antisymmetric pair features.

A subtle point to consider is that the coupled angular basis elements~\eqref{eq:hamiltonian-coupled} transform as $Y^\mu_\lambda$ under $SO(3)$ (proper) rotations, but behave differently under $O(3)$ (improper) rotations. 
While spherical harmonics transform under inversion as polar tensors $\hat{i}\rep|\lambda\mu>=(-1)^\lambda\rep|\lambda\mu>$, some of the coupled basis terms transform as pseudotensors, $\hat{i}\rep|\lambda\mu>=(-1)^{\lambda+1}\rep|\lambda\mu>$. 
For instance, consider the case of $(\tl=1,\tl*=1)$. The product $\rep|1 \tm>\otimes\rep|1\tm*>$ is even under inversion (because each term is odd); and so the $\lambda=1$ coupled terms $\rep|\tl=1;\tl*=1;\lambda=1\ \mu>$  must also be even - and thus behave as a pseudovector. 
In general, the parity of a block of the Hamiltonian in the coupled form $\rep<\hat{H}; \tnnlammu||A_{ii'}>$ is $\sigma=(-1)^{\tl+\tl*+\lambda}$. 
Given that the NICE iterations~\eqref{eq:nice-body}-\eqref{eq:nice-center} provide a natural strategy to track the parity of the equivariant features, it is possible to exploit this additional symmetry, that has been shown to increase the transferability of the resulting models in the context of $\lambda$-SOAP-based symmetry-adapted Gaussian process regression\cite{gris+19book}. 
We conclude noting that for the matrix elements corresponding to the same element and orbital index, parity and index exchange symmetry are linked, resulting in some blocks of the coupled-momentum Hamiltonian being identically zero.

\subsection{Symmetry-adapted regression}

Armed with a set of symmetry-adapted features, and with a transformation of the Hamiltonian into irreducible symmetry blocks, we can proceed to construct regression models. 
Given the emphasis we give to the construction of well-principled features, we restrict ourselves to simple models, but it would be relatively simple to build an $O(3)$ equivariant neural network\cite{will+19jcp,ande+19nips}, using the equivariant features as inputs. 

For each block $\qblock =(\e\tnl; \e'\tnl*; \eta)$ (where $\eta$ indicates the equivariance and symmetry with respect to index exchange, if relevant), and rotationally-equivariant component $\lambda$,  a linear regression model reads
\begin{multline}
\rep<\hat{H}; \qblock ; \lambda \mu|| A_{ii'}>
\approx \\
\sum_q \rep<\hat{H}; \qblock ; \lambda;||q> \rep<q||A_{ii'}; \operatorname{sym}(\qblock );  \lambda \mu>,\label{eq:sa-ridge-model}
\end{multline}
where we indicate with ``$\operatorname{sym}(\qblock )$'' that the features should have symmetries matching those of the Hamiltonian block. Even though it may be advantageous to separately tune the hyperparameters of the representation depending on the type of block, here we use the same features to regress all blocks having the same symmetry. 
The symmetry-adapted regression weights $w^{(\qblock ,\lambda)}_q = \rep<\hat{H}; \qblock ; \lambda;||q> $ can be determined computing a symmetrized covariance matrix
\begin{equation}
C_{qq'}^{(\qblock ,\lambda)} = \!\!\! \sum_{ii'\in \qblock , \mu}\!\!\!
\rep<A_{ii'}; \operatorname{sym}(\qblock );  \lambda \mu||q>
\rep<q'||A_{ii'}; \operatorname{sym}(\qblock );  \lambda \mu>
\end{equation}
and feature-weighted targets
\begin{equation}
z_q^{(\qblock ,\lambda)} = \sum_{ii'\in \qblock , \mu}\!\!\! \rep<\hat{H}; \qblock ; \lambda \mu||A_{ii'}>\rep<A_{ii'}; \operatorname{sym}(\qblock );  \lambda \mu || q>,
\end{equation}
and then evaluating a ridge-regression expression
\begin{equation}
 \mbf{w}^{(\qblock ,\lambda)} = (\mbf{C}^{(\qblock ,\lambda)} + \sigma^2 \mbf{1})^{-1} \mbf{z}^{(\qblock ,\lambda)},
\end{equation}
where $\sigma$ is a regularization parameter, that can be optimized by cross-validation. 

We also test a kernel-based symmetry-adapted Gaussian-process regression (SA-GPR) model. We compute symmetry-adapted kernels that generalize to the multi-center case the $\lambda$-SOAP kernel of Ref.~\citenum{gris+18prl}
\begin{multline}
\krn^{(\qblock ,\lambda)}_{\mu\mu'}(A_{i i_2},A'_{i' i'_2})=
\sum_q
\rep<A_{ii_2}; \operatorname{sym}(\qblock );  \lambda \mu || q>\\\rep<q||A'_{i'i_2'}; \operatorname{sym}(\qblock );  \lambda \mu'>\label{eq:sa-gpr-kernel}
\end{multline}
and then formulate a SA-GPR ansatz
\begin{equation}
\rep<\hat{H}; \qblock ; \lambda \mu|| A_{ii'}>
\approx \sum_{M\mu'} b^{(\qblock ,\lambda)}_{M\mu'} 
\krn^{(\qblock ,\lambda)}_{\mu\mu'}(A_{i i'},M),
\end{equation}
where $M$ indicates a collection of ``active'' points selected among the atomic environments (or pairs) in the training set.
In the projected-process approximation, one can then compute the kernel regression weights by 
\begin{equation}
\mbf{b}^{(\qblock ,\lambda)} = (\bK_{NM}^T\bK_{NM}+\sigma^2 \bK_{MM})^{-1}\bK_{NM}^T
\end{equation}
where $\bK_{MM}$ and $\bK_{NM}$ indicate respectively the matrices collecting the kernels between the active-set clusters and those between training and active-set clusters\cite{deri+21cr}. 
The cluster index and the angular momentum index $\mu$ are merged in a single index.
An important difference with previous applications of SA-GPR is that one should be careful when using non-linear kernels built by combining an invariant $\lambda=0$ and a $\lambda>0$ kernel\cite{will+19jcp,wilk+19pnas}.
In fact, some of the invariant features may be zero because of molecular symmetries, while some of the corresponding $\lambda>0$ features are not. %
This leads to scenarios in which $\krn^0_{00}$, but not $\krn^\lambda_{\mu\mu'}$,  vanishes.
Thus, we define non-linear kernels as products of the equivariant linear kernel\eqref{eq:sa-gpr-kernel} with a polynomial in the invariant kernel,
\begin{equation}
\krn^{(\boldsymbol{\zeta}, \qblock ,\lambda)}_{\mu\mu'}(A,A') = \krn^{( \qblock ,\lambda)}_{\mu\mu'}(A,A')
\sum_{p=0}^{p_\text{max}} \zeta_p \krn^{( \qblock ,0)}(A,A')^p.
\end{equation}
The $p=0$ term ensures a non-zero value also in the few cases in which the scalar kernel vanishes because of symmetry.

\section{Continuous symmetries and molecular symmetries}

By incorporating all permutation, rotation and inversion symmetries, a Hamiltonian predicted based on these features is guaranteed to be equivariant to all point-group operations that may be present for a symmetric molecular geometry. %
This is a consequence of the fact that point groups are a subgroup of the combination of $O(3)$ operations and permutations,\cite{longuet1963symmetry,wigner1959group, bunker2006molecular} -- an observation that also generalizes to space groups symmetries, thanks to the fact that $N$-center representations are also translationally invariant.
As a consequence, the eigenfunctions of $\Hhat$ must transform according to irreducible representations of the point group, and those that transform into each other under point group operations must be degenerate. 

\begin{figure}
    \centering
    \includegraphics[width=1.0\linewidth]{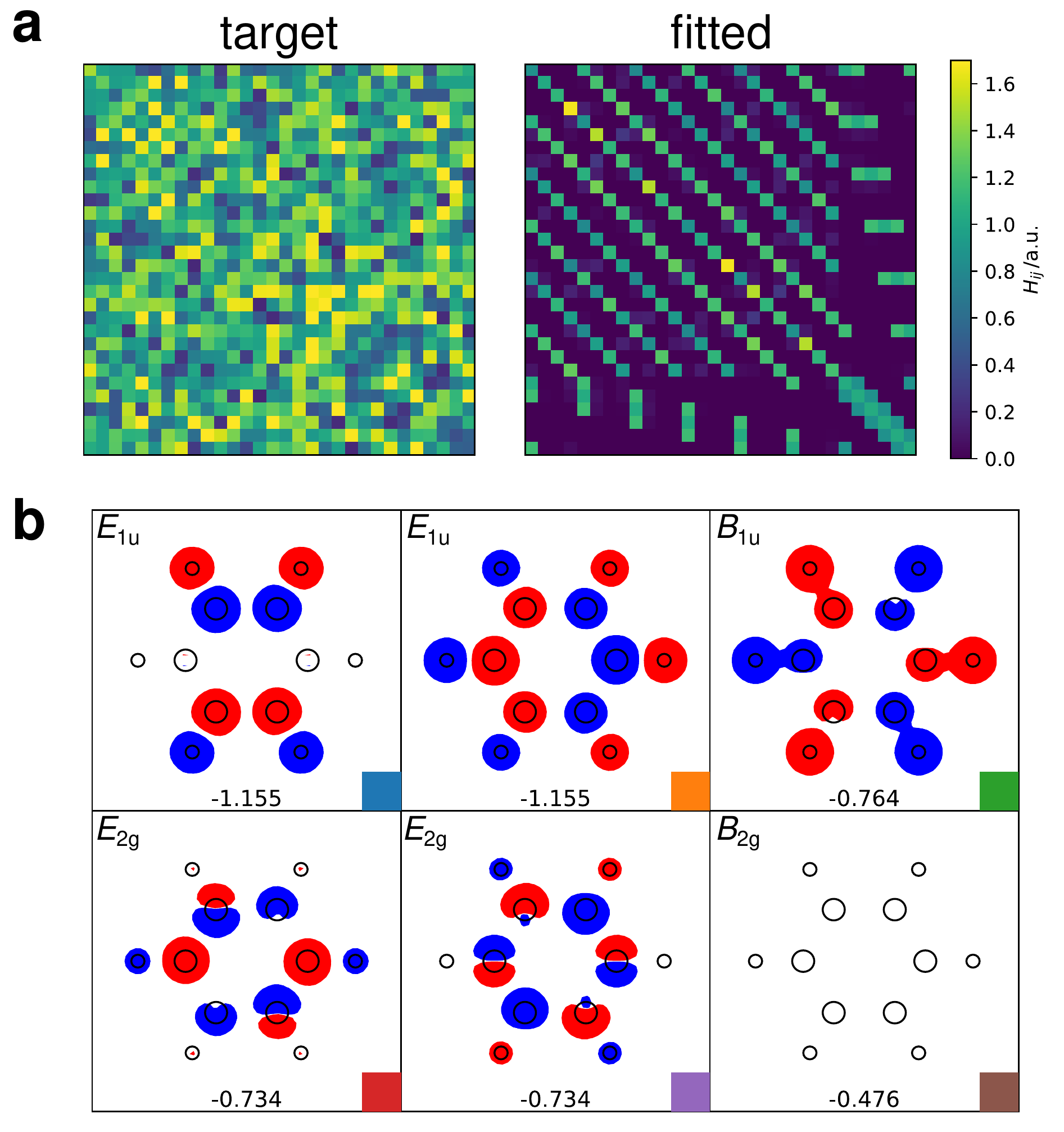}
    \caption{(a) A minimal-basis Hamiltonian for benzene. Left: filled with random numbers, right: predicted by a symmetry-adapted model after learning on the random target. (b) The first six eigenstates of the predicted Hamiltonians. This is just a cartoon representation built from the isocontours of a fictional wavefunction built from atomic orbitals that combine a single Gaussian function centered on the atoms with the appropriate real spherical harmonics, evaluated in the plane of the molecule. The sixth state has a nodal surface in the plane.  }
    \label{fig:benzene-eigv}
\end{figure}
\newcommand{\dsixh}{\ensuremath{\text{D}_\text{6h}}}
To explore this observation, consider benzene at its equilibrium geometry. Atomic orbitals centred on each of the atoms generate a reducible representation $\Gamma$ of $\dsixh$, the point group that the geometry corresponds to. This reducible representation can be decomposed uniquely into irreducible components,
\begin{equation}
  \Gamma = a_{1} \Gamma_{1} \oplus a_{2} \Gamma_{2} \oplus \dots \oplus a_{k} \Gamma_{k},
\end{equation}
where $\Gamma_{l}$ are irreducible representations of $\dsixh$. 
Simply by recognising that the Hamiltonian satisfies the point group symmetries (i.e., commutes with all the symmetry elements), we can conclude that the molecular orbitals obtained from this basis must transform according to the irreducible representations appearing on the right-hand side of the decomposition written above. 
This has an immediate consequence on the structure of the energy levels: if there are $k$ $l$-dimensional irreducible representations in the decomposition, there must be $k$ energy levels which are $l$-fold degenerate. 
To further refine the energy level diagram, we can estimate the coupling through the Hamiltonian of symmetry-adapted linear combinations of atomic orbitals. The coupling between symmetry-adapted linear combinations belonging to different irreducible representation is strictly zero. 
The magnitude of the other couplings cannot be determined based on symmetry, but can be estimated from physical principles.

\begin{figure}
    \centering
    \includegraphics[width=1.0\linewidth]{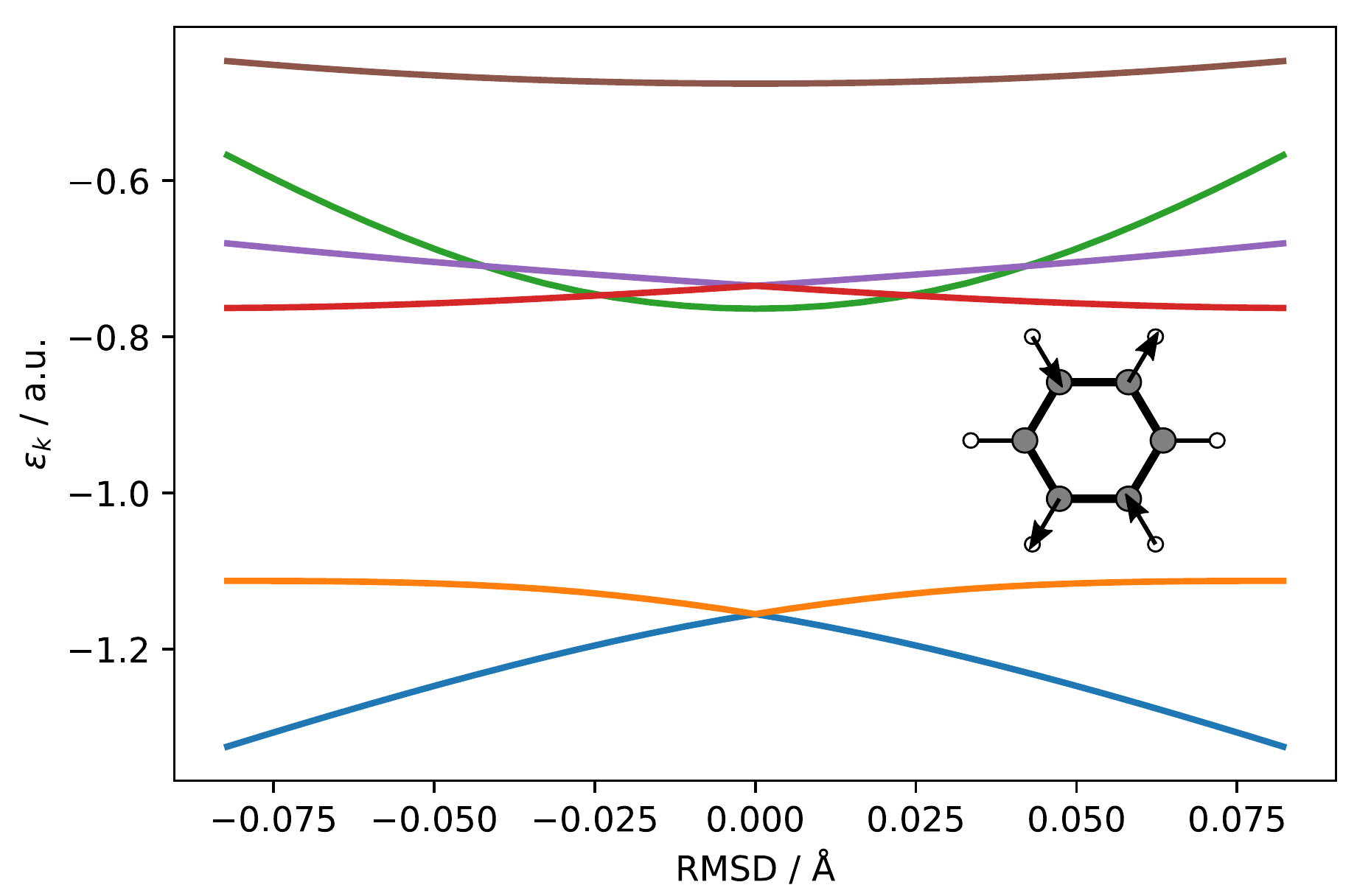}
    \caption{An example of the changes in energy levels associated with a $E_{2g}$ symmetric deformation of the benzene molecule, computed from a minimal-basis Hamiltonian. The matrix elements are predicted from a model trained on a single random-valued matrix. \rev{The energy levels are color-coded following the same key as in Fig.~\ref{fig:benzene-eigv}.}  }
    \label{fig:benzene-disp}
\end{figure}

The procedure just described is familiar to all chemists as molecular orbital theory, a cornerstone of modern chemistry, which can be remarkably successful in describing the bonding and stability of molecules. Exploiting point-group symmetry is an indispensable part of the procedure, and the small number of simple rules that are commonly used to refine an energy level diagram are independent of the constraints imposed by symmetry. 
In other words, modifying the rules cannot affect the features of the energy level structure that are constrained on symmetry grounds.
Since the Hamiltonians generated by our model are guaranteed to satisfy point group symmetries, the model is forced to operate in much the same way. The model extracts from training data the information that cannot be determined based on symmetry alone, and it cannot help but predict the correct energy level structure and molecular orbital symmetries.

As a numerical example, we build a $30\times 30$ matrix filled with uniform random numbers between 0 and 1, symmetrized by adding its own transpose, and we interpret it as the single-electron Hamiltonian of benzene at the equilibrium geometry in a minimal valence basis (C$2s$, C$2p$, H$1s$). We train our model on this single configuration. 
The model cannot reproduce the random Hamiltonian because it is incompatible with the molecular symmetry.  Instead, it learns a matrix which reveals the symmetry of the system (Figure~\ref{fig:benzene-eigv}a). 
Furthermore, the molecular orbitals of the predicted Hamiltonian (which we compute assuming the functions in the minimal basis are orthogonal) transform according to irreducible representations of $\dsixh$, and the degeneracy of the energy levels is thus correct. 
The representation generated by the minimal basis decomposes into irreducible components as follows,
\begin{equation}
  \Gamma = 2 \textrm{A}_{1\textrm{g}} \oplus \textrm{A}_{2u} \oplus 2 \textrm{B}_{1\textrm{u}}
  \oplus \textrm{B}_{2\textrm{g}} \oplus \textrm{E}_{1\textrm{g}} \oplus
  2 \textrm{E}_{1\textrm{u}} \oplus 2 \textrm{E}_{2\textrm{g}} \oplus
  \textrm{E}_{2\textrm{u}}.
\end{equation}
The predicted Hamiltonian thus exhibits at least six doubly-degenerate energy levels (corresponding to molecular orbitals with E symmetry).  Figure~\ref{fig:benzene-eigv}b shows the six lowest-lying molecular orbitals, labelled with irreducible representations, four of which are doubly degenerate. Even though the model is trained with nonsensical data, the appearance of the predicted molecular orbitals and their degeneracies are qualitatively correct. Quantitative accuracy could be obtained by exposing the model to physically-meaningful training data. 
Note that exactly the same model, exposed to no more training data, would predict a Hamiltonian of e.g. Buckminsterfullerene, satisfying all of the 120 point group symmetries -- although it might also fulfil additional symmetries, because $\dsixh$ is not a subgroup of $\text{I}_\text{h}$.

This numerical example demonstrates the well-known fact that symmetry considerations are often indispensable when solving problems in quantum mechanics. 
Any behaviour governed by point group symmetry must be replicated by our model. Consider for instance  the breaking of degeneracy as the molecule distorts away from a high-symmetry geometry, as in the celebrated Jahn-Teller theorem.\cite{jahn1937stability} Continuing with the numerical example,  the symmetric squares of $\textrm{E}_{1\textrm{u}}$ and $\textrm{E}_{2\textrm{g}}$ are the same,
\begin{equation}
  \left[ \textrm{E}_{1\textrm{u}} \otimes \textrm{E}_{1\textrm{u}} \right] = \textrm{A}_{1\textrm{g}} \oplus
  \textrm{E}_{2\textrm{g}}.
\end{equation}
Therefore, we expect the degeneracy of the $\textrm{E}_{1\textrm{u}}$ and $\textrm{E}_{2\textrm{g}}$ energy levels of our predicted Hamiltonian to be broken by distortions of the geometry along normal modes of $\textrm{E}_{2\textrm{g}}$ symmetry. The variation of the degenerate levels along such a normal mode should vary to first order in the displacement. On the other hand, the $\textrm{B}_{1\textrm{g}}$ and $\textrm{B}_{1\textrm{u}}$ energy levels should not vary to first order in the displacement. This analysis is confirmed by the variation of the energy levels in Figure~\ref{fig:benzene-disp} and has a clear consequence on the way that incorporating symmetry facilitates learning: the symmetry of a geometry not only places constraints on the Hamiltonian at that geometry but also dictates how the Hamiltonian must change when the geometry is distorted. This imposes a subtle structure on the form on the Hamiltonian as a function of geometry that a symmetry-unaware model would have to learn through exposure to training data.

\rev{
\subsection{Broken symmetries}

Even though in most cases molecular orbitals obey the symmetry imposed by the point-group symmetry of the nuclei, in the presence of excited states - and more in general for states with a non-zero total angular momentum of the wavefunction - the symmetry of the electron density can be broken, leading in principle to lower symmetry also for the self-consistent effective Hamiltonian. Even the density of isolated atoms, when prepared in a state with angular momentum greater than zero, is not spherically-symmetric\cite{kohn-fert00pra}. 
Another - simpler - example involves the case in which an external field is present that lowers the symmetry of the system.
This scenario can be easily incorporated in this symmetry-adapted scheme. Take for instance the linear regression model in Eq.~\eqref{eq:sa-ridge-model}, and consider the case where an electric field is included in the construction of the Hamiltonian. The electric field couples only to $\lambda=1$ components of the Hamiltonian, and so \emph{only for $\lambda=1$} models one should allow for weights that depend on the $\mu$ index
\begin{multline}
\rep<\hat{H}; \qblock ; 1 \mu|| A_{ii'}>
\approx \\
\sum_q \rep<\hat{H}; \qblock ; \mu||q> \rep<q||A_{ii'}; \operatorname{sym}(\qblock ); 1 \mu>.
\end{multline}
Note that this approach would require fitting a separate model for each value of the symmetry-breaking field; an alternative would be building a symmetry-adapted model of the \emph{response} of the matrix elements of the Hamiltonian, that can be written in a way that is equivariant with respect to rotations of the reference frame, similar to how dipole moments and polarizabilities (the first and second-order response of the scalar energy to an electric field) can be learned in a fully equivariant fashion.\cite{wilk+19pnas,veit+20jcp} 
}

\rev{
\section{Role of the orbital basis}

The difficulty in learning the matrix elements of the Hamiltonian depends substantially on the atomic basis -- obviously in terms of the number of models that must be built, but also in terms of the learning rate, and the range of interatomic correlations that have to be included in the pair features. Here we discuss some of the approaches that can be used to simplify or make more efficient the learning problem, focusing in particular on the common case in which the Hamiltonian is not the final goal, but only an intermediate quantity in evaluating its eigenvalue spectrum. 

\subsection{Regression of non-orthogonal Hamiltonians} \label{sub:lowdin}

The atomic orbital basis is non-orthogonal, which means that -- in order to obtain the single-particle wavefunction coefficients matrix $\mbf{U}$ and eigenvalues $\boldsymbol{\epsilon}$ from the Hamiltonian matrix $\mbf{H}$ -- one needs to solve a generalized eigenvalue equation, 
\begin{equation}
\mbf{H} \mbf{U} = \mbf{S} \mbf{U}\operatorname{diag}\boldsymbol{\epsilon},
\label{eq:eigv-gen}
\end{equation}
where $\mbf{S}$ is the overlap matrix of the basis functions. 
Building a separate model for the overlap matrix is often unnecessary, because for most families of atomic orbitals the entries in $\mbf{S}$ can be computed analytically and with minimal  effort. 
It may however be advantageous from a ML perspective to target a matrix that can be diagonalized directly to obtain the same single-particle energy levels as from Eq.~\eqref{eq:eigv-gen}. This is because $\mbf{S}$ is often ill-conditioned, leading to a magnification of errors when solving~\eqref{eq:eigv-gen} -- similar to what is observed in the case of models of the electron density\cite{gris+19acscs}.

An orthogonal Hamiltonian $\bar{\mbf{H}}$  can be obtained by computing the L\"owdin-orthogonalization of $\mbf{H}$,
\begin{equation}
\bar{\mbf{H}} = \mbf{S}^{-1/2} \mbf{H} \mbf{S}^{-1/2}.
\end{equation}
\rev{It might appear that this transformation would disrupt the symmetries of $\mbf{H}$; in fact, if  $\mbf{S}^{-1/2}$ is the symmetric inverse square root of the overlap, it does transform precisely as $\mbf{H}$ under rotations or permutations of the atom indices. }
Thus, the orthogonal Hamiltonian $\bar{\mbf{H}}$ obeys exactly the same symmetries as $\mbf{H}$. Even though, as we shall see, it is more difficult to learn, it leads to  comparably-accurate predictions for the eigenvalues. Unless otherwise specified, in all numerical benchmarks we will build models for the L\"owdin-orthogonalized Hamiltonian matrix.

\begin{figure}
    \centering
    \includegraphics[width=1.0\linewidth]{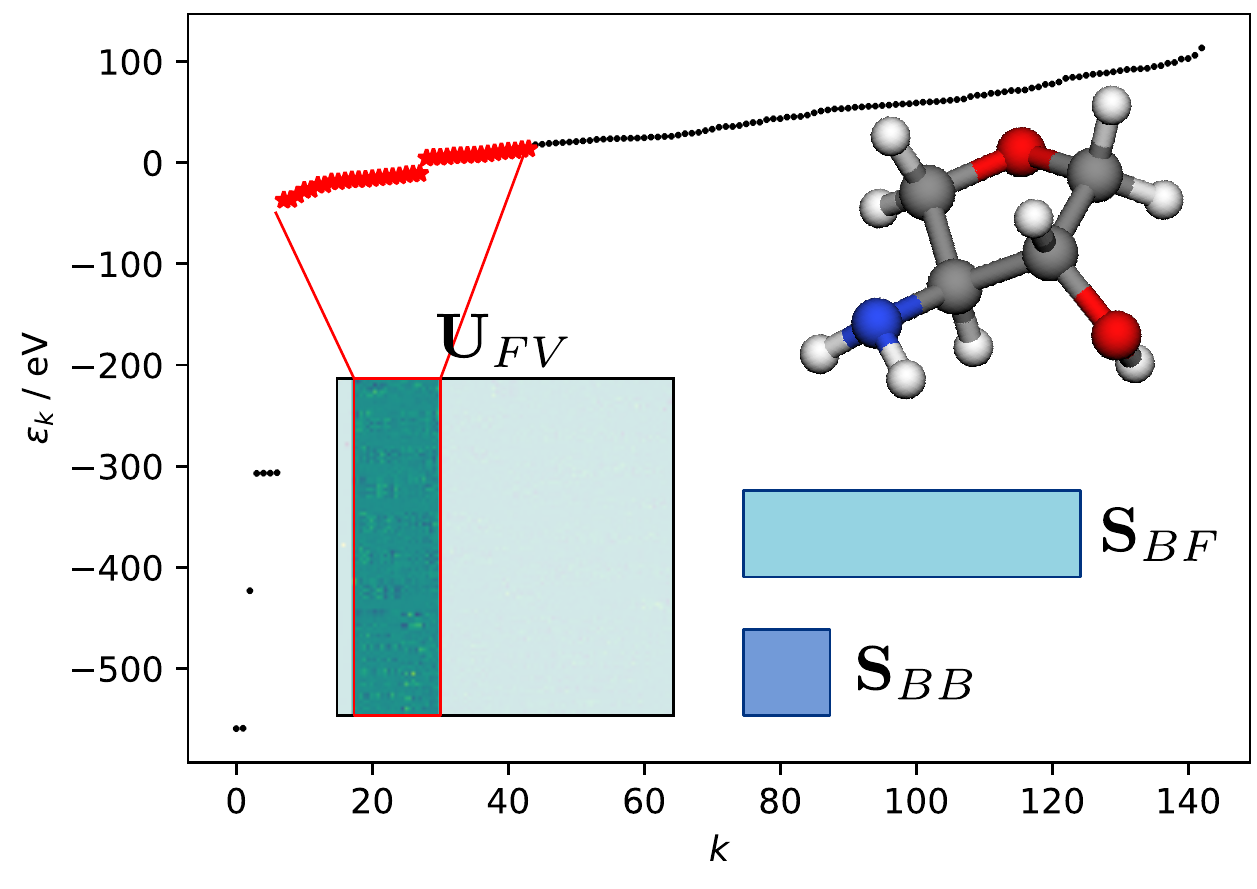}
    \caption{A schematic depiction of the construction of a symmetry-adapted projected Hamiltonian. The most relevant section of the eigenvalue spectrum of a large-basis Hamiltonian is selected (red stars) and the corresponding eigenvectors projected on a minimal auxiliary basis using the overlap matrix $\mbf{S}_{MF}$. The minimal basis Hamiltonian can then be built to reproduce exactly the target spectrum, as discussed in the text.   }
    \label{fig:saph-scheme}
\end{figure}

\subsection{Symmetry-adapted projected Hamiltonian}
\label{sec:SAPH}
One of the challenges inherent in the construction of machine-learning models of $\ncent$-centers properties is that the number of items grows very rapidly, making the training and prediction more time-consuming. 
In the case of matrices associated with an atomic-orbitals representation of a quantum operator, this problem is made worse by the need of an atomic basis of fairly large size. 
As we shall see, and as is observed in Refs.~\citenum{west-maur21cs,gast+20jcp}, using a large basis is also detrimental when learning the Hamiltonian because it introduces a large number of virtual states that are delocalized, physically meaningless, and difficult to learn. 
A possible strategy, similar to that followed in Ref.~\citenum{west-maur21cs}, is to learn a small pseudo-Hamiltonian whose eigenvalues are compared with the subset of the single-particle energy levels that are relevant for the application at hand. 
Our equivariant models would allow doing so in a symmetry-adapted way, and is likely to bring large improvements compared to the prediction of an invariant pseudo-Hamiltonian.

An alternative approach, followed in the construction of maximally-localized Wannier orbitals\cite{marz-vand97prb,marz+12rmp},  of quasi-atomic minimal basis orbitals\cite{qian+08prb} that have also been applied to simplify the construction of SchnOrb-type deep-learning models\cite{gast+20jcp}, is to determine a small basis with desirable properties (locality, good conditioning, ...) and use it to write an effective Hamiltonian that reproduces the most useful part of the eigenvalue spectrum.
To preserve exactly the symmetry properties of the target, the only requirement is that the minimal basis is atom-centered and corresponds to well-defined irreps of $O(3)$. Therefore, we project explicitly the relevant molecular orbitals on a minimal atomic basis, obtaining a symmetry-adapted projected Hamiltonian (SAPH), that has the correct symmetries and the same eigenvalues as the full-basis Hamiltonian. This allows us to discuss some of the subtleties that are associated with targeting the eigenvalues rather than the Hamiltonian. The approach is also of practical utility, as it reduces both effort and error when one is only interested in a portion of the eigenspectrum of $\Hhat$. 

We compute the generalized eigenvalue decomposition of the full Hamiltonian, and select a subset $V$ of its eigenstates. We label $\mbf{U}_{FV}$ the block of the eigenvector matrix associated with the selected set, and $\boldsymbol{\epsilon}_V$ the subset of corresponding eigenvalues. 
For instance, one may drop core levels, and high-energy virtual orbitals (Fig.~\ref{fig:saph-scheme}).
We then choose a minimal basis $B$ of the same overall size as $V$. In this work we pick from the same basis used for the full calculation a number of atomic orbitals corresponding to the valence states, but one could also take a completely different basis, as long as it has well-defined symmetry behavior: finding an optimal choice is an interesting problem of itself, as it will likely combine issues of locality and of conditioning of the overlap matrix.\cite{schu-vand18jctc} 
The overlap matrix between minimal and full basis, $\mbf{S}_{BF}$ can be used to find the projection $\mbf{P}_{BV}=\mbf{S}_{BF}\mbf{U}_{FV}$. This projection is non-orthogonal, but an orthogonalized version can be obtained by first performing a singular value decomposition, $\mbf{P}_{BV} = \mbf{U}_P \boldsymbol{\Sigma}_P \mbf{V}_P^T$ and then computing  $\bar{\mbf{P}}_{BV} = \mbf{U}_P \mbf{V}_P^T$.

At this point, there are important symmetry considerations to be made. For a symmetric molecule, the minimal basis spans a collection of irreducible components $\Gamma_B$, and the selected molecular orbitals $\mbf{U}_{FV}$ a possibly different collection $\Gamma_V$. 
If the  two don't match, the minimal basis projection may have lower rank than $B$, making it impossible to approximate a non-singular Hamiltonian. This problem can be mitigated by keeping track of the orthogonalized projections of the molecular orbitals starting from the low-energy valence states. High-energy (empty) states that cannot be projected on the orthogonalized basis can then be replaced with even higher-energy states, until one obtains a valence set that has a non-singular projection on the minimal basis. 
These problems are a manifestation of the general issue of tracking the character of eigenstates when there is a quasi-continuum of eigenvalues. Molecular orbitals  can swap in and out of the selected set, leading to discontinuities in the construction of  $\mbf{P}_{BV}$ which can be avoided by choosing a block of orbitals that is clearly separated from the remainder (e.g. dropping core levels is usually not a problem), but that affect negatively the performance of the ML model at the high end of the spectrum, where usually there is no clear gap to exploit. 
The problem shares many formal and practical similarities with the use of maximally-localized Wannier orbitals in interpolating single-particle energy bands in solids as a function of electron momentum $\mathbf{k}$. The symmetry considerations, the problem of bands changing character with changing $\mathbf{k}$, as well as that of dealing with virtual states that blend into a continuum\cite{souz+01prb,vita+20npjcm} are analogous to the issues we observe here when changing the atomic structure and composition. 

Once a suitable projected matrix has been determined, we can proceed to build a minimal-basis SAPH. Here two options are available:
\begin{equation}
\bar{\mbf{H}}_{BB} = \bar{\mbf{P}}_{BV} \operatorname{diag}\boldsymbol{\epsilon}_V \bar{\mbf{P}}_{BV}^T 
\end{equation}
which is an orthogonal Hamiltonian that yields the exact large-basis eigenvalues $\boldsymbol{\epsilon}_V$ when solving a standard eigenvalue problem, and 
\begin{equation}
\mbf{H}_{BB} = \mbf{S}_{BB}^{1/2}\bar{\mbf{P}}_{BV} \operatorname{diag}\boldsymbol{\epsilon}_V \bar{\mbf{P}}_{BV}^T \mbf{S}_{BB}^{1/2}
\end{equation}
that yields the correct eigenvalues when solving a generalized eigenvalue problem $\mbf{H}_{BB} \mbf{U} = \mbf{S}_{BB}\mbf{U}\operatorname{diag}\boldsymbol{\epsilon}_V  $.
Similar to the case of the full Hamiltonian we find that the entries of the non-orthogonal SAPH are easier to learn, but in our preliminary tests the error on the predicted eigenvalues to are comparable or higher than those obtained from the orthogonal SAPH predictions. 
This is however one of the many aspects of this study that deserve a more thorough investigation in the future. 
}

\section{Examples and benchmarks}

We now discuss results for a few examples of increasing complexity and diversity. Our main objective will be to demonstrate the advantages of using symmetry-adapted features, but we will also discuss the role of the model, and compute a few of the properties one may want to extract from the exercise of predicting an atomic Hamiltonian. 
We note also that existing schemes show wildly different performances depending on the basis set used, and the number of energy levels targeted by the model. Ref.~\citenum{west-maur21cs} provides a thorough comparison, showing that errors for the same dataset and model vary by more than an order of magnitude depending on the range of the spectrum targeted. 
\rev{
We compare calculations targeting the full Hamiltonian matrix and reduced models based on the SAPH construction, that show differences highlighting the subtle interplay between basis set and model accuracy. 
}
We do not attempt to optimize the hyperparameters of the descriptors (cutoff distance, Gaussian smearing, ...) -- in part for simplicity, in part because we use the same features for all blocks, and we want to avoid optimizing them for a specific type of matrix elements. 
The only optimization we consider, for each block type and train set size, is that of the ridge regularization, which we perform by grid search and 3-fold cross validation.

\begin{figure*}[tbp]
    \centering
    \includegraphics[width=0.8\linewidth]{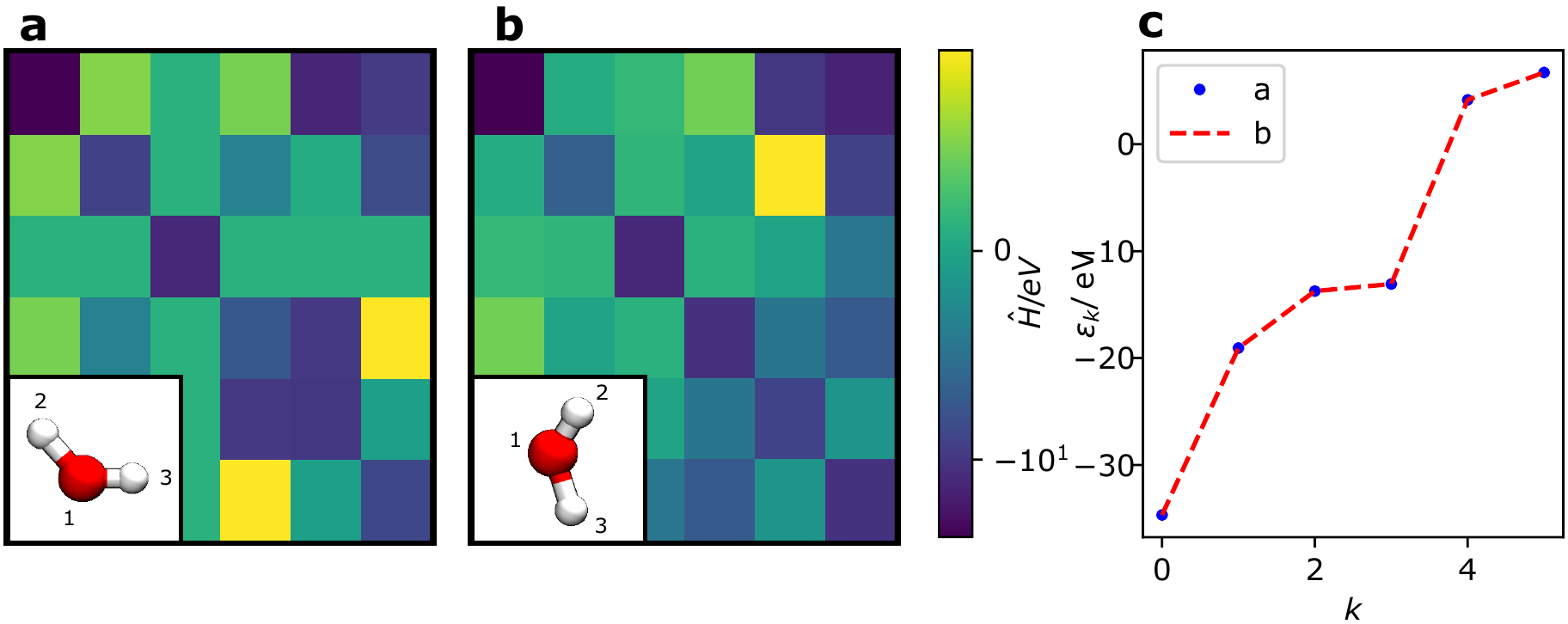}        
    \caption{Equivariance of predictions under rotation of a distorted water molecule and permutation of its \ce{H} atoms. The prediction in Fig b) is obtained by a model trained on the orthogonalized SAPH matrix of a molecule in a). Even though the individual matrix elements are different, the eigenvalues in c) are predicted to be exactly the same, within machine precision. }
    \label{fig:compare-rotperm-water}
\end{figure*}

\subsection{Datasets, computational details,  and metrics}
We used three different datasets: (1) a dataset of 1000 \ce{H2O} configurations that was originally introduced in Ref.~\citenum{gris+18prl} to demonstrate symmetry-adapted regression of tensors, that can be downloaded from a public data record\cite{matcloud18c}; (2) the \ce{CH3CH2OH} trajectory used in Ref.~\citenum{schu+19nc}, which we use together with the corresponding  Hamiltonian matrices to provide a 1:1 comparison; (3) the subset of 6868 molecules from the QM7b dataset\cite{mont+13njp} that contains only \ce{CHNO} atoms. 

For datasets (1) and (3) we performed restricted Hartree Fock calculations using the quantum chemistry code PySCF\cite{sun2020recent,sun2018pyscf}, \rev{using a  def2-SVP basis for all atoms. }%
A convergence threshold of 1e-10 atomic units for both energies and gradients was set.
The Hamiltonian and the overlap matrices obtained in the atomic-orbital basis were then transformed to the L\"owdin orthogonalized basis as described in Section~\ref{sub:lowdin} and an orthogonalized symmetry adapted Hamiltonian with a minimal basis (H1$s$ and O2$s$, O2$p$ orbitals) were obtained as described in Section~\ref{sec:SAPH}.

We define the mean square error associated with the overall prediction $\tilde{H}$ of the (orthogonal) Hamiltonian matrices over a given test set as 
\begin{multline}
\text{MSE}_\text{full} = \frac{1}{N_{\text{test}}} \sum_{A\in \text{test}} \frac{1}{N_A}
\\\times \sum_{i\qblock  i'\qblock '}
\lvert \rep<\hat{H}; \qblock ; \qblock '||A_{ii'}>
- \rep<\tilde{H}; \qblock ; \qblock '||A_{ii'}> \rvert^2,
\label{eq:mse-full}
\end{multline}
where $N_{\text{test}}$ is the number of structures in the test set and $N_A$ indicates the size of the Hamiltonian matrix, i.e. the total number of atomic orbitals, and we use the notation of Eq.~\eqref{eq:hamiltonian-decoupled} to indicate summation over all the elements in the Hamiltonian matrix.
We also compute errors for individual symmetry-adapted blocks,
\begin{multline}
\text{MSE}_{\qblock \lambda} = \frac{1}{N_{\text{test}}} \sum_{A\in \text{test}}  \frac{1}{N_{A}} \\
\times \sum_{\mu, ii'\in \qblock } \lvert \rep<\hat{H}; \qblock ; \lambda\mu||A_{ii'}>
- \rep<\tilde{H}; \qblock ; \lambda\mu||A_{ii'}> \rvert^2,
\label{eq:mse-blocks}
\end{multline}
where we use the coupled-blocks notation~\eqref{eq:hamiltonian-coupled} in which $\qblock \equiv (\tnl;\tnl*;\pm)$. Eq.~\eqref{eq:mse-blocks} is defined in such a way that summing over all blocks (making sure to account for the multiplicity of each block in the full Hamiltonian) yields the same value as the full MSE~\eqref{eq:mse-full}.
We also compute errors for the eigenspectrum
\begin{equation}
\text{MSE}_\epsilon = \frac{1}{N_{\text{test}}} \sum_{A\in\text{test}}\frac{1}{N_A} \sum_k \lvert \epsilon_k - \tilde{\epsilon}_k \rvert^2
\end{equation}
where $N_{\text{test}}$ is the number of structures in the test set and $\epsilon_k$ and $\tilde{\epsilon}_k$ are the sorted eigenvalues of the Hamiltonian and their predictions. 
We typically report root mean square errors (RMSE) that are simply the square roots of the definitions given above, or mean absolute errors (MAE) defined by summing over the absolute value rather than the square of the errors. 

\begin{figure}[tb]
    \centering
    \includegraphics[width=1.0\linewidth]{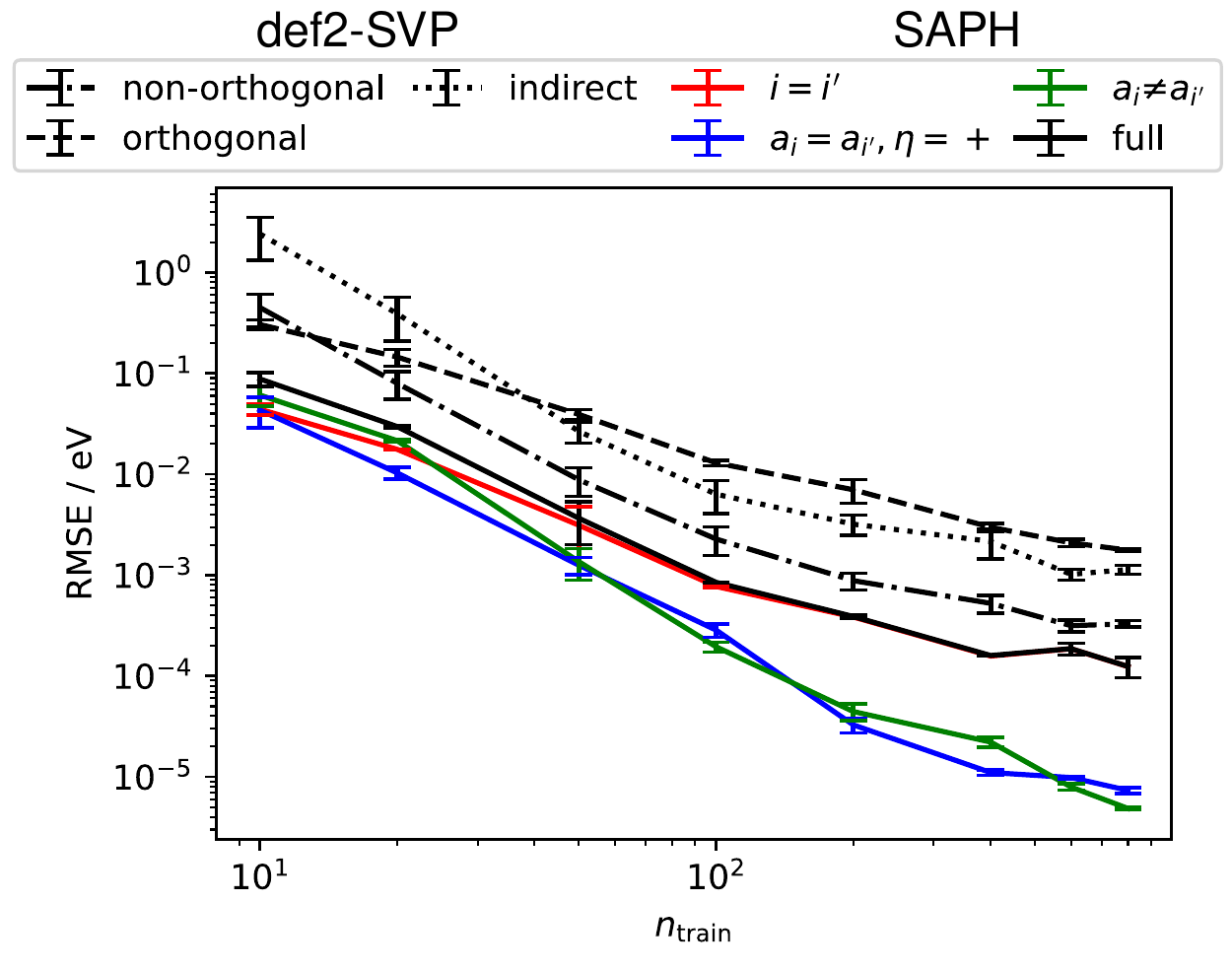}
    \caption{Learning curve for the matrix elements in different blocks of the Hamiltonian for a test set of 200 water molecules. Full lines correspond to linear ridge regression models trained on the orthogonalized SAPH matrices obtained from the reference calculations in the def2-SVP basis, based on $(\ncent=2, \nu=1)$ features. 
    Errors, in the def2-SVP basis, for predicting both the full orthogonal ($\bar{\mbf{H}}$) and non-orthogonalized $\mbf{H}$ matrices are shown with dashed and dash-dot lines respectively, while dotted lines indicates the error for computing $\bar{\mbf{H}}$ by orthogonalizing the prediction of $\mbf{H}$. }
    \label{fig:water-lc}
\end{figure}

\subsection{Water molecule}

We begin showing results for the prediction of the valence Fock matrix of a dataset of 1000 distorted water molecules. Given that water is composed of only three atoms, $\nu=2$ features provide a complete basis to regress the diagonal blocks,  and $\ncent=2, \nu=1$ features for the off-diagonal terms. 
We then limit ourselves to a linear regression scheme, using features built from neighbor density expansion coefficients  with $\rcut=4$~\AA, $\sigma_a=0.3$~\AA, $\nmax=12$, $\lmax=8$. The radial basis was further optimized using the scheme of Ref.~\citenum{gosc+21jcp}.  
Figure~\ref{fig:compare-rotperm-water} demonstrates the equivariance of the model by training a linear model on the orthogonal SAPH  of the Fock matrix for a selected distorted structure, and then making a prediction for another that is obtained by rotating the molecule and swapping the indices of the H atoms. The matrix elements differ, but in a precisely equivariant way: the eigenvalues of the two matrices are identical within machine precision. 

\begin{figure}[tb]
    \centering
    \includegraphics[width=1.0\linewidth]{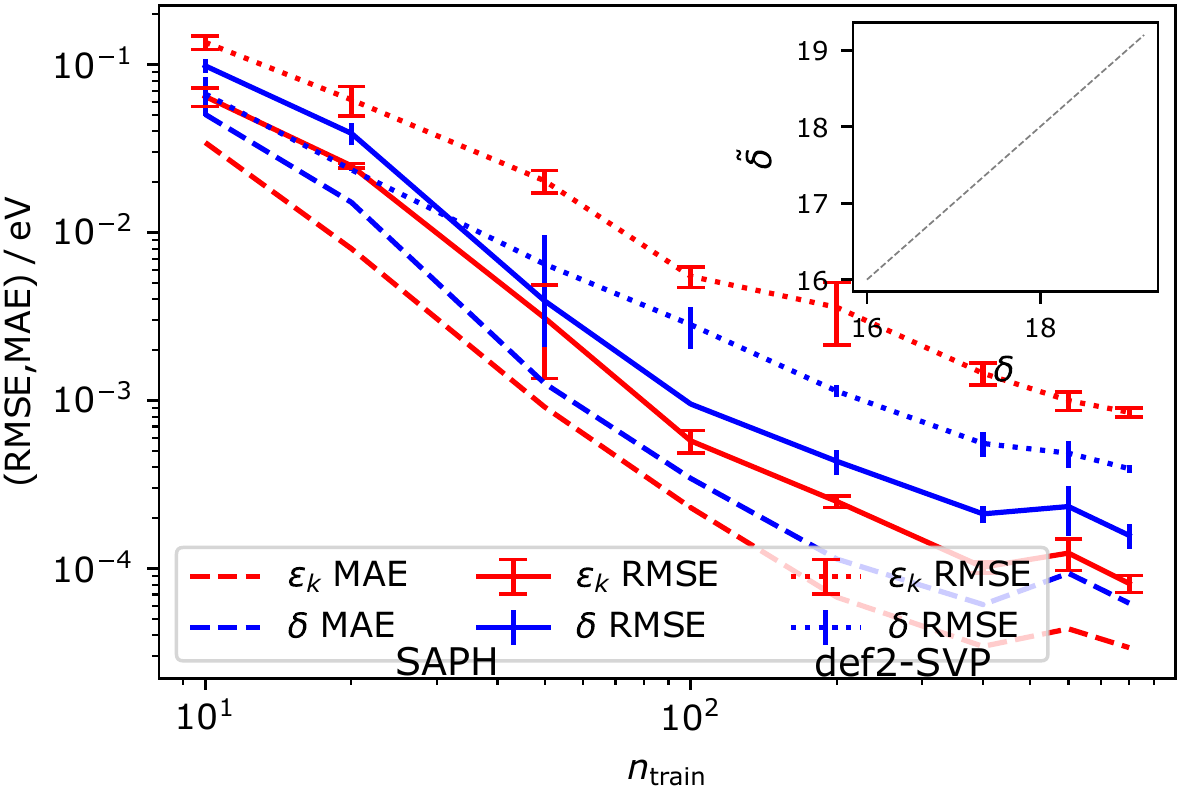}
    \caption{Learning curves for the errors in the eigenvalue spectrum $\epsilon_k$ (red) and the HOMO-LUMO gap $\delta$ (blue) for a test set of 200 water molecules. Full and dashed curves correspond to RMSE and MAE, evaluated for the models trained on the orthogonalized SAPH in Fig.~\ref{fig:water-lc}. The inset shows a parity plot for the predicted versus reference values of $\delta$ for $n_{\text{train}}=800$. 
    Dotted lines show the corresponding errors computed for the orthogonalized def2-SVP Hamiltonian $\bar{\mbf{H}}$}
    \label{fig:water-gap}
\end{figure}

We then consider the training of a symmetry-adapted ridge regression model on up to 800 structures, assessing the accuracy of the predictions on up to 200 configurations.
Learning curves for the def2-SVP Hamiltonian show little sign of saturation, despite using a simple linear model, and achieve a $\text{RMSE}_\text{full}$ well below 10~meV with fewer than 100 configurations.  
Fig.~\ref{fig:water-lc} compares the accuracy of predicting the non-orthogonal Hamiltonian $\mbf{H}$, the L\"owdin-orthogonalized $\bar{\mbf{H} }$, and that of estimating the orthogonalized Hamiltonian by applying the exact $\mbf{S}^{-1/2}$ to the non-orthogonal prediction. Note the error for predicting $\mbf{H}$ is almost 10 times smaller than that for modeling $\bar{\mbf{H} }$, but that -- even without considering the inconvenience and the further errors connected with predicting or computing the overlap matrix -- orthogonalizing the predicted $\mbf{H}$ increases the error fourfold. 
\rev{
We also evaluate the errors for predicting the orthogonal SAPH that reproduces the four valence states, and two empty states, achieving errors that are at least two times smaller than for the full Hamiltonian.
The largest errors are associated with the diagonal blocks (that constitute by far the largest fraction of the SAPH matrix); note that the $\eta=-$ off-diagonal blocks are exactly zero. 

Steady convergence of the learning curves is also observed for the eigenvalues (Fig.~\ref{fig:water-gap}), with a more substantial improvement in accuracy for RMSE$_\epsilon$ when using the SAPH to restrict the eigenvalue spectrum to the most relevant states. 
Even though a precise comparison is not possible given the differences in the details of the calculations, the accuracy of predictions, which show an error at (or well below) 1 meV compare very favorably with recent equivariant NN results\cite{unke2021se3equivariant} (RMSE$_\epsilon$ of 2meV with 500 training structures),  as well as with SchnOrb\cite{schu+19nc} (RMSE$_\epsilon$ of 7meV).
}

\subsection{Ethanol}

We then consider the case of a larger molecule, for which $(\ncent+\nu)=3$ features cannot provide a full linear basis to learn the matrix elements of the Hamiltonian.
We take a random subselection of up to 4500 Hamiltonian matrices and corresponding structures from the HF/def2-SVP calculations for ethanol in Ref.\citenum{schu+19nc}, and further subdivide it into a train set of up to 4000 structures and a test set of 500 structures. 
\rev{We construct SAPHs using a subset of the  def2-SVP basis functions corresponding to a minimal atomic basis. }
We build $(\ncent=1,\nu=1)$, $(\ncent=1,\nu=2)$, $(\ncent=2,\nu=1)$ and $(\ncent=2,\nu=2)$ features starting from neighbor density expansion coefficients  with $\rcut=5$~\AA, $\sigma_a=0.3$~\AA, $\nmax=12$, $\lmax=8$, with a data-optimal radial basis and an iterative PCA contraction akin to the NICE scheme to reduce the number of $(\ncent=2,\nu=2)$ features.

\begin{figure}[tb]
    \centering
    \includegraphics[width=1.0\linewidth]{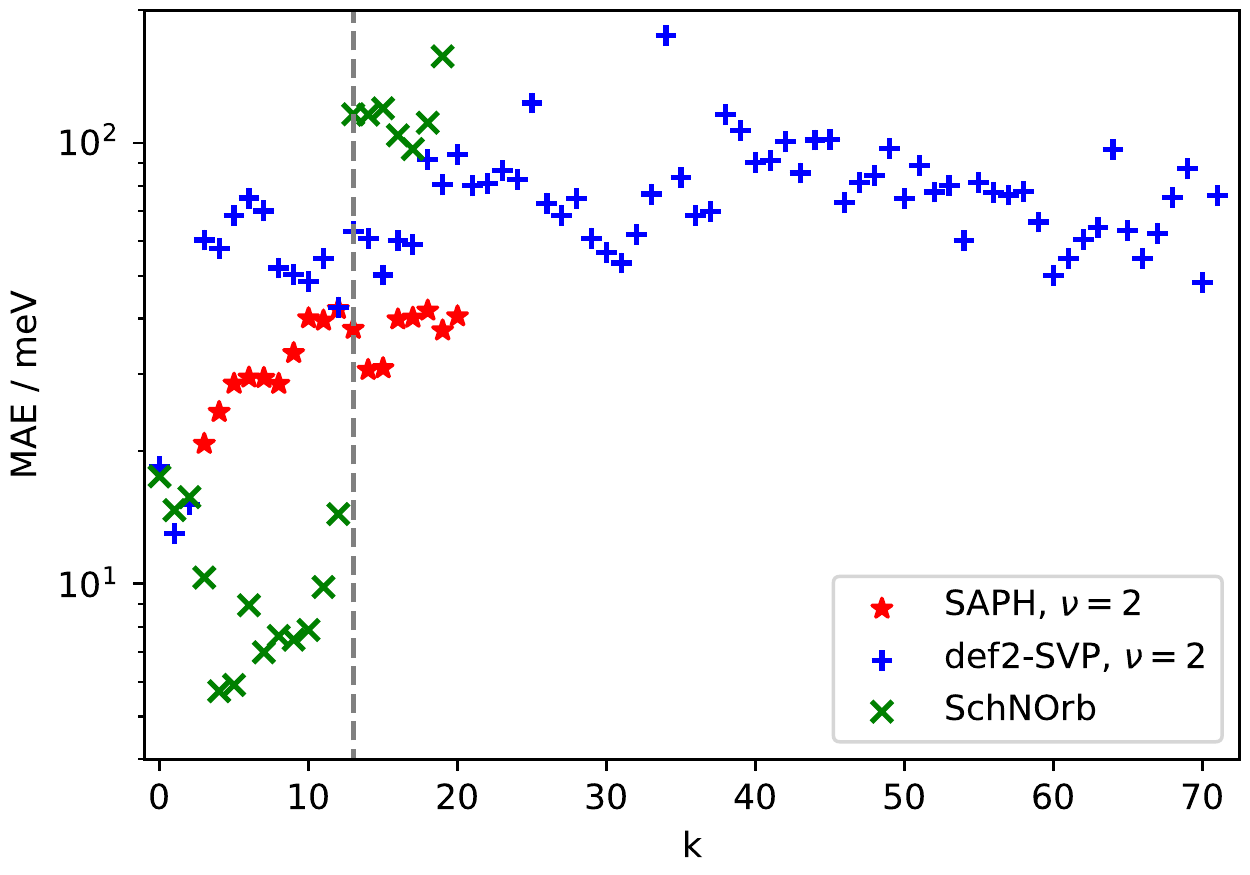}
    \caption{MAE for individual Hamiltonian eigenvalues $\epsilon_k$. Results for the $\nu=2$ linear model, with 900 training structures for the def2-SVP Hamiltonian and 4000 training points the SAPH, is compared with the results from SchNOrb\cite{schu+19nc}, with 25k training structures. }
    \label{fig:eigenval-schnorb}
\end{figure}

\begin{figure}
    \centering
    \includegraphics[width=1.0\linewidth]{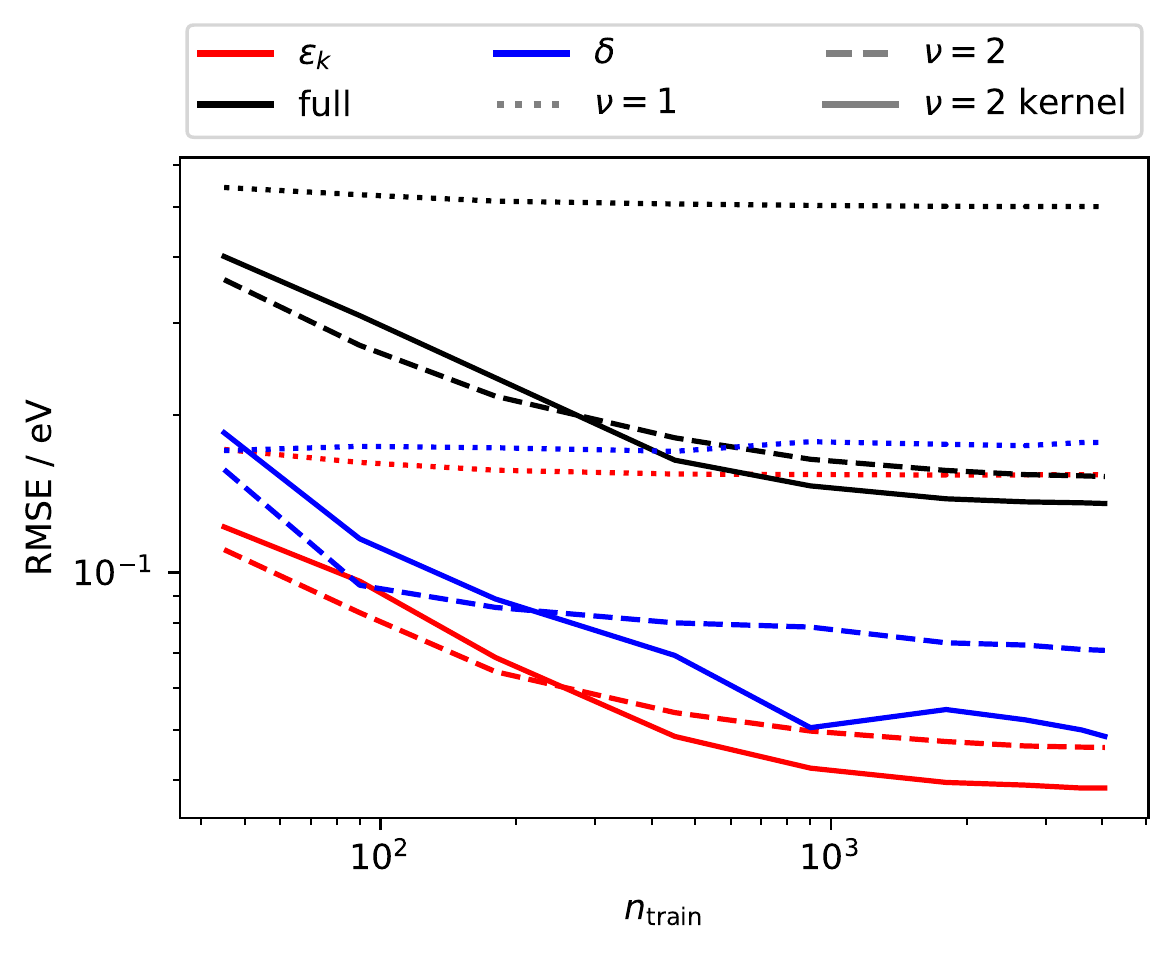}
    \caption{RMSE$_\text{full}$, RMSE$_\epsilon$ and  RMSE$_\delta$ on the SAPH for 100 \ce{CH3CH2OH} molecules from the same dataset as in Ref.\citenum{schu+19nc}. Different curves correspond to linear models using $\nu=1,2$ pair features, and to a SA-GPR model based on $\nu=2$ features. }
    \label{fig:lc-full-eigen-gap-ethanol}
\end{figure}

\rev{
Linear models based on $\nu=2$ features computed for the full orthogonal def2-SVP Hamiltonian show errors that are much larger than for water, around 600meV for $\text{RMSE}_\text{full}$ and 100meV for $\text{RMSE}_\epsilon$. The error is not too large when compared with the range covered by the full spectrum, that spans a range of more than 100 eV, but it is much larger than what is quoted for SchnOrb and PhisNet. 
There are at least two reasons for the lower performance. First, we target the matrix elements of the full Hamiltonian: as shown in Fig.~\ref{fig:eigenval-schnorb}, our linear model yields an error that is roughly constant throughout the spectrum, while SchnOrb has a much lower error on the valence states, but a much higher one on the unoccupied states. 
We can partly address this issue by using an SAPH target, that indeed leads to a reduction of the error down to approximately 40 meV $\text{RMSE}_\epsilon$, 130 meV $\text{RMSE}_\text{full}$, and 40 meV for the HOMO-LUMO gap. It will be interesting to compare these results with models that specifically target the eigenspectrum while being based on a simple linear ansatz for the matrix elements. 

On a more fundamental level, $\nu=2$ representations, that provide a full linear basis to express the properties of a triatomic molecule, are insufficient to describe the structural correlations found in a larger molecule. 
This is seen clearly from the learning curves in Fig.~\ref{fig:lc-full-eigen-gap-ethanol}: $\nu=1$ features lead to almost immediate saturation of the model performance, but also the $\nu=2$ models saturate almost completely at $n_\text{train}=1000$.
While it would be interesting to compare the performance of deep learning models in the data poor regime (the only results available in the literature are trained on 25'000 configurations), the inability of improving the accuracy by increasing the train set size is problematic. 

Previous studies have shown that the construction of non-linear kernel models lead to dramatic improvement of the performance for both scalar\cite{deri+21cr} and tensorial\cite{wilk+19pnas}  extensive properties. 
Here, on the other hand, using a non-linear kernel yields only marginal improvements of performance, and does not eliminate the saturation of the learning curves. 
Investigating the reasons for the difference in behavior when targeting atom and pair properties is left for future studies, as well as increasing the correlation order of the representations, which is conceptually trivial but requires a considerable implementation effort.
}

\begin{figure}[tbp]
    \centering
    \includegraphics[width=1.0\linewidth]{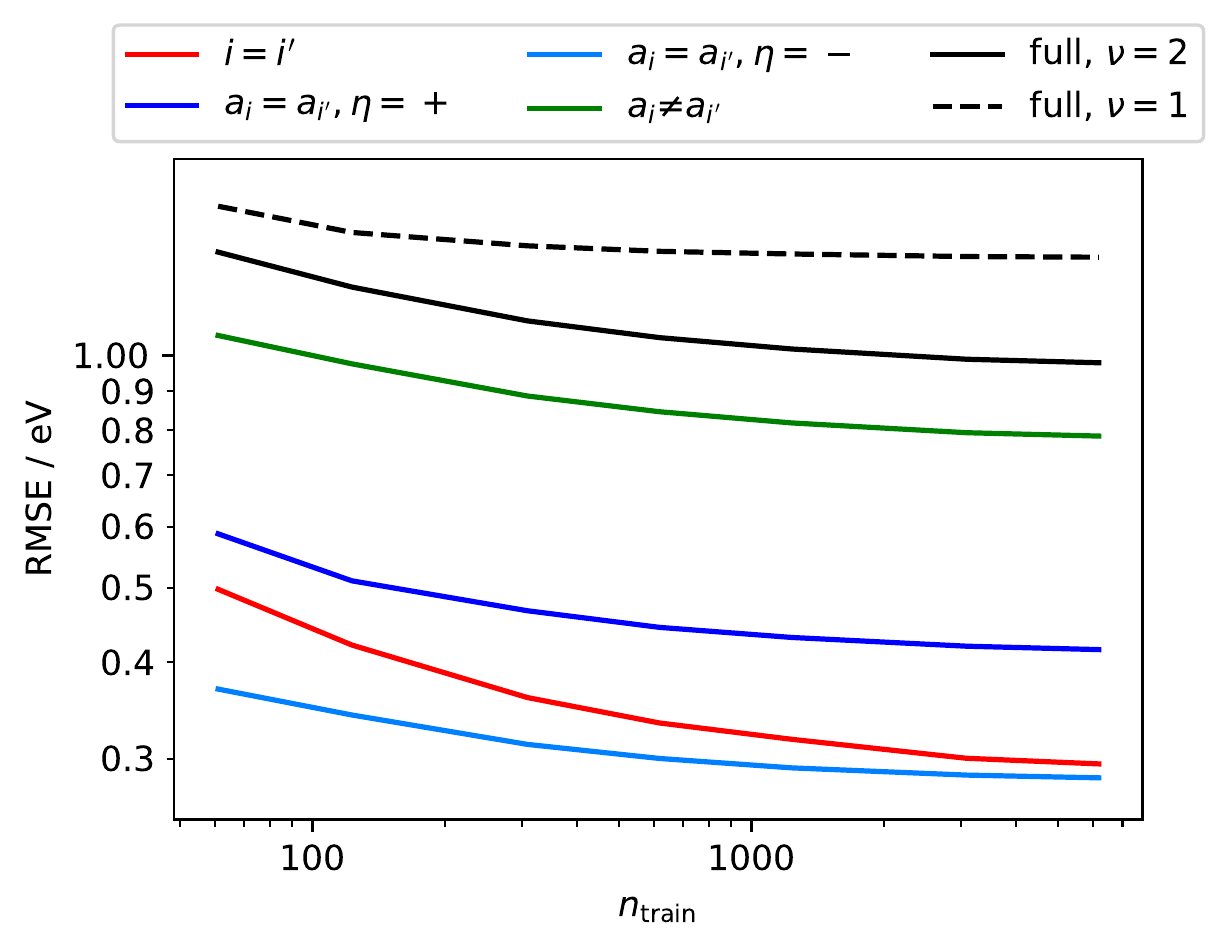}
    \caption{Learning curves for the different blocks of the Fock matrix for the QM7-CHNO dataset. Full and dashed lines correspond to a linear regression and a sparse kernel regression model.  }
    \label{fig:qm7-lc}
\end{figure}

\subsection{Organic molecules}

Finally, we turn to the more challenging task of learning the Hamiltonian of a diverse database of organic molecules, containing up to 7 \ce{CNO} atoms, with different degrees of H saturation.\cite{mont+13njp}
Even with a small SVP valence basis, this is a task of great complexity, with almost 400 models that have to be trained for the different types of blocks. 
We thus use a SAPH construction so that the SVP eigenvalues are projected on a minimal atomic valence basis, which reduces by an order of magnitude the number of blocks.
We choose heuristically a long cutoff $\rcut=7$~\AA, $\sigma_a=0.3$~\AA, and the radial scaling\cite{will+18pccp} $r_0=2$~\AA, $m=2$, and generate an optimal radial basis following the procedure in Ref.~\citenum{gosc+21jcp}.
The large number of different species leads to an explosion of the raw number of features: for the relatively parsimonious density discretization parameters $\nmax=8, \lmax=6$ that we use, the raw number of $(N=2, \nu=2)$ equivariant features would be in excess of $10^9$ per pair. 
Thus, it becomes essential to perform a data-driven selection, which we implement following the same iterative PCA scheme used for the NICE features in Ref.~\citenum{niga+20jcp}.

\begin{figure}[tbp]
    \centering
    \includegraphics[width=1.0\linewidth]{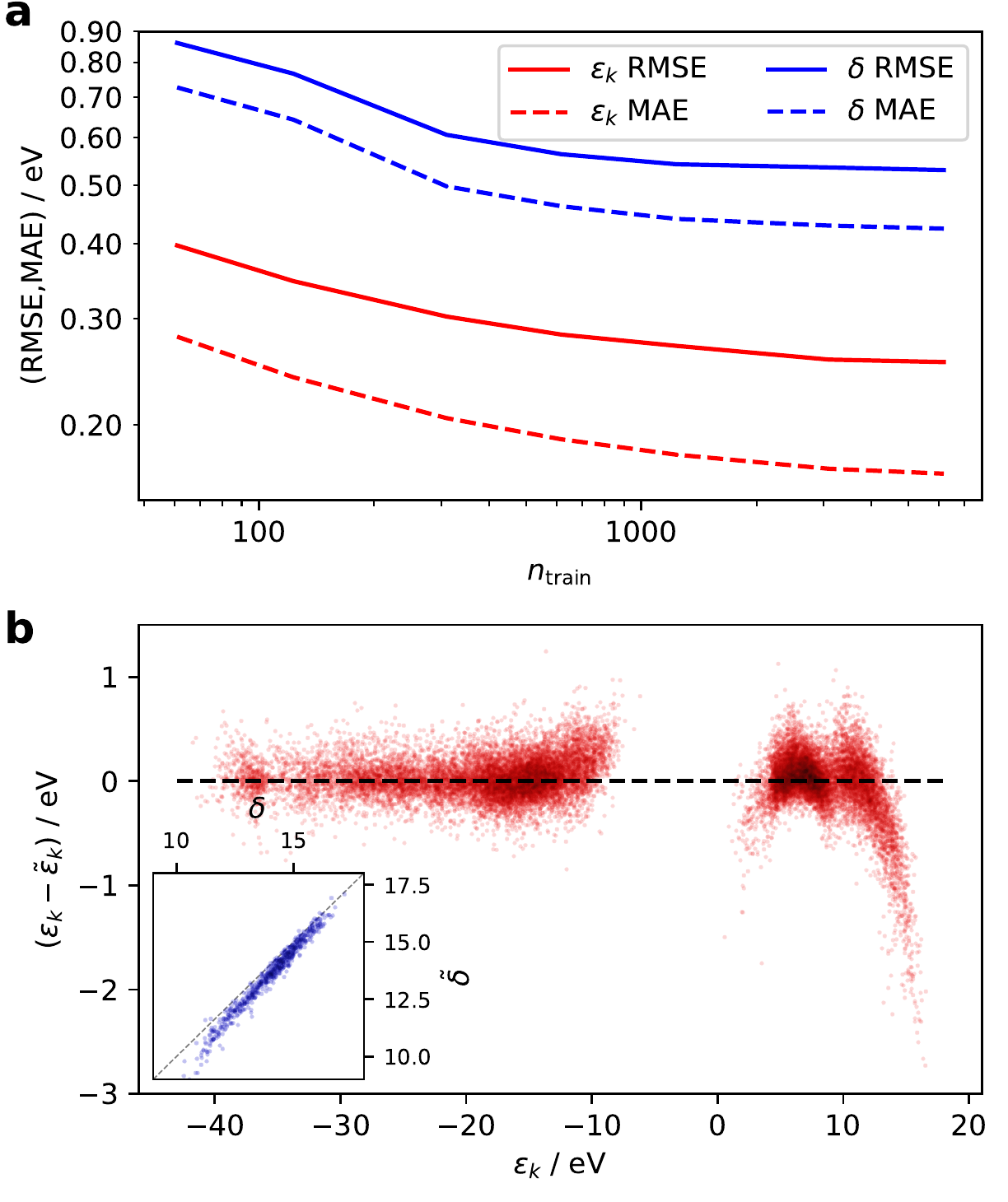}
    \caption{Errors in the prediction of the single-particle energy levels of molecules from the QM7-CHNO dataset. (a) Learning curves for the mean eigenvalue error, and for the error in the HOMO-LUMO gap $\delta$. (b) Signed error in the prediction of electronic eigenvalues $\epsilon_k$ as a function of the energy level. Note the larger spread for states close to the HOMO-LUMO gap (which is around 0~eV) and for the upper end of the projected virtual orbitals. The inset shows the parity plot for the predicted gap $\delta$. }
    \label{fig:qm7-eva}
\end{figure}

As shown in Fig.~\ref{fig:qm7-lc}, even a rudimentary linear model (that only describes the dependence of the matrix elements on 3-body neighbor correlations) achieves rapidly an error that is only two times larger than what we observed  for the much simpler case of \ce{CH3CH2OH}. 
It shows however clear signs of saturation, with all the blocks -- in particular those corresponding to off-diagonal terms -- reaching an almost flat learning curve profile. 
Moving to a non-linear, $\zeta_p=\{1,0.2,0.04\}$ model delays saturation for the diagonal blocks, but leads to a slight degradation of performance for the off-diagonal terms (see SI). %
\rev{The errors for predictions of the eigenspectrum are larger than for ethanol, with a mean RMSE of 0.3~eV (0.15eV MAE) at the largest train fraction of 90\% of the QM7-CHNO dataset. The main issue appears to be related to a saturation of the learning curves, which points at the excessive simplicity of the regression models as a key limitation. 
The error distribution as a function of energy is rather uneven (Fig.~\ref{fig:qm7-eva}b), with systematic deviations observed for the states close to the gap. The error in the HOMO-LUMO gap is 0.53 eV RMSE (0.43 eV MAE), with larger errors for the structures with lower gap and a systematic underestimation. 
It is also interesting to observe the large, systematic deviations for the states at the high end of the spectrum, that confirm the difficulty in learning virtual states, and suggest that the discontinuities that occur in the construction of the SAPH have a direct effect on the model accuracy.

Besides improving the models and the construction of a minimal orbital basis, there is a more pragmatic approach that could be followed to immediately improve the accuracy and transferability of ML schemes to learn effective Hamiltonians. 
The construction of the initial guess for the Hamiltonian based on atomic orbitals is usually much less time consuming than achieving self-consistency. Thus, one can predict just the difference between the self-consistent effective Hamiltonian and the initial guess. For QM7-CHO, doing so results in a substantial reduction of the error, with an overall eigenvalue RMSE of 0.14~eV (0.09eV MAE) and a more uniform error distribution around the gap (HOMO-LUMO errors being 0.14 eV RMSE, 0.10 eV MAE) and for the virtual states (see SI). 
}

\section{Conclusions}

We have introduced a construction of symmetry-adapted, atom-permutation and rotation equivariant representations that are suitable to describe quantities associated with multiple atomic centers. 
The construction is closely related to the density-correlation features that have been used to describe atom-centered environments, and that underlie the vast majority of machine-learning frameworks for the microscopic properties of matter. 
We propose a practical implementation of the general scheme, that is based on the NICE construction,\cite{niga+20jcp} but several existing frameworks, including atom-centered symmetry functions,\cite{behl11jcp} the atomic cluster expansion,\cite{drau20prb} or moment tensor potentials\cite{shap16mms} can be extended along the same lines. It is important to stress that, even if we only show examples for molecular systems, these $N$-center representations can also be readily applied to the condensed phase. 

We then present as an application the construction of machine-learning models of an effective \rev{single-particle} Hamiltonian, written in an atom-centered basis. Formulating the problem in a fully symmetry-adapted fashion requires manipulating the entries in the Hamiltonian to separate them into blocks with a well-defined behavior with respect to rotation, inversion, particle exchange and Hermitian symmetry.
The reward for building a symmetry-adapted model is that, on top of the general symmetries that are explicitly built in, it also automatically incorporates the molecular orbital rules associated with point-group symmetries, when present. 
We give a striking demonstration of this property by training a model on a random-valued Hamiltonian for a benzene molecule, showing that the \emph{predicted} Hamiltonian yields orbital symmetries that are fully consistent with the expected $\dsixh$ group characters. 
\rev{ We address the steep increase in model complexity that would arise from the use of large basis sets -- that are needed to achieve converged quantum calculations -- by constructing a symmetry-adapted projected Hamiltonian that replicates with a minimal basis the eigenstates of a more converged calculation.
}

We then benchmark the method on problems of increasing complexity, and find that symmetry-adapted features provide excellent accuracy for a homogeneous dataset of distorted \ce{H2O} molecules -- with linear regression achieving an accuracy comparable to non-symmetry-adapted deep learning models with a fraction of the training set size.  
\rev{ The case of ethanol highlights the importance of focusing the models on the most relevant part of the spectrum, which can be seen by comparing the model accuracy for the Hamiltonian expressed in a typical quantum chemical basis set, and that for a symmetry-adapted effective Hamiltonian that only reproduces the valence and low-lying unoccupied states. 
The lack of features with higher neighbor correlation order results in early saturation of the learning curves, which is not eliminated by using non-linear kernel models.  A similar saturation is seen also for a dataset of small organic molecules,  where the problem is exacerbated by the heterogeneity of the atomic environments. }

\rev{
Further work is needed to fully elucidate the interplay between the choice of basis, the hyperparameters of the representation, and the accuracy of the model. Targeting of a subset of the energy levels, as done in Ref.~\citenum{west-maur21cs}, while using a symmetry-adapted model for the intermediate prediction of an effective Hamiltonian, is a very promising research direction. 
}
From the point of view of building fully equivariant descriptors of $N$-center atomic clusters, the next step will be to introduce higher-body-order terms, with $\nu>2$, either explicitly or through more sophisticated non-linear models.
The general construction we present here provides an easily-extendable framework to do so, as well as to tackle the modeling of 3-center integrals, and higher-$N$ quantities, bringing the full set of ingredients of quantum chemistry calculations within the reach of equivariant machine learning schemes. 

\section*{Data availability}

The archived source code used to perform the calculations presented in this work is available from Ref.~\citenum{ncenter-zenodo}. Commented datafiles that can be used to reproduce these results can be dowloaded from Ref.~\citenum{matcloud21c}.

\section*{Supplementary Material}
The supplementary material contains additional derivations and benchmarks that ccomplement and support the results presented in the main text.

\begin{acknowledgments}
JN and MC acknowledge support by the NCCR MARVEL, funded by the Swiss National Science Foundation (SNSF), the Swiss National Science Foundation (Project No. 200021-182057), and the European Research Council (ERC) under the European Union’s Horizon 2020 research and innovation programme (grant agreement No 101001890-FIAMMA) . We would like to thank Andrea Grisafi, Max Veit, Susi Lehtola, Reinhard Maurer and Julia Westermayr for helpful insights and stimulating discussions. 

\end{acknowledgments}

\onecolumngrid
\clearpage
\newpage

\end{document}